\documentclass[a4paper,fleqn,usenatbib]{mnras}

\usepackage{mathptmx}

\usepackage[T1]{fontenc}
\usepackage{ae,aecompl}

\usepackage{graphicx}	
\usepackage{amsmath}	
\usepackage{amssymb}	

\usepackage{listings} 
\usepackage{threeparttable}

\title[The dusty pinwheel WR 98a]{Pinwheels in the sky, with dust: 3D modeling of the Wolf-Rayet 98a environment}

\author[T. Hendrix et al.]{
Tom Hendrix$^{1}$,
Rony Keppens$^{1}$\thanks{Corresponding Author, E-mail: rony.keppens@wis.kuleuven.be},
Allard Jan van Marle$^{1}$,
Peter Camps$^{2}$,
\newauthor
Maarten Baes$^{2}$,
and Zakaria Meliani$^{3}$
\\
$^{1}$Centre for mathematical Plasma Astrophysics, Department of Mathematics, KU Leuven, Celestijnenlaan 200B, 3001 Leuven, Belgium\\
$^{2}$Sterrenkundig Observatorium, Universiteit Gent, Krijgslaan 281, B-9000 Gent, Belgium\\
$^{3}$Observatoire de Paris, 5 place Jules Janssen 92195 Meudon, France
}

\date{Accepted XXX. Received YYY; in original form ZZZ}

\pubyear{2015}

\begin{document}
\label{firstpage}
\pagerange{\pageref{firstpage}--\pageref{lastpage}}
\maketitle

\begin{abstract}
The Wolf-Rayet 98a (WR 98a) system is a prime target for interferometric surveys,
since its identification as a ``rotating pinwheel nebulae", where infrared images display a spiral dust lane revolving with a 1.4 year periodicity. WR 98a hosts a WC9+OB star, and the presence of dust is puzzling given the extreme luminosities of Wolf-Rayet stars.
We present 3D hydrodynamic models for WR 98a, where dust creation and redistribution are self-consistently incorporated. Our grid-adaptive simulations resolve details in the wind collision region at scales below one percent of the orbital separation ($\sim$ 4 AU), while simulating up to 1300 AU.
We cover several orbital periods under conditions where the gas component alone behaves adiabatic, or is subject to effective radiative cooling. In the adiabatic case, mixing between stellar winds is effective in a well-defined spiral pattern, where optimal conditions for dust creation are met.
When radiative cooling is incorporated, the interaction gets dominated by thermal instabilities along the wind collision region, and dust concentrates in clumps and filaments in a volume-filling fashion, so WR 98a must obey close to adiabatic evolutions to demonstrate the rotating pinwheel structure.
We mimic Keck, ALMA or future E-ELT observations and confront photometric long-term monitoring. We predict an asymmetry in the dust distribution between leading and trailing edge of the spiral, show that ALMA and E-ELT would be able to detect fine-structure in the spiral indicative of Kelvin-Helmholtz development, and confirm the variation in photometry due to the orientation. Historic Keck images are reproduced, but their resolution is insufficient to detect the details we predict.
\end{abstract}

\begin{keywords}
hydrodynamics -- radiative transfer -- methods: numerical -- stars: Wolf-Rayet -- binaries: general -- infrared: stars
\end{keywords}



\section{Introduction}
\subsection{Wolf-Rayet stars and the presence of dust}
Wolf-Rayet (WR) stars, named after their discoverers Charles Wolf and Georges Rayet in 1867, have unusually broad emission lines, with peaks located in a pattern similar to the (hydrogen) Balmer series. Only much after their discovery, it was realised that these lines were caused by helium instead, and it became clear that these stars come in two main varieties: those with strong lines of helium and nitrogen (WN subtype) and those with strong helium and carbon lines (WC subtype). Even more recently a third WO subtype with oxygen lines was added. This WO subtype is much rarer than the two others: currently only nine WO stars are known \citep{2015A&A...581A.110T}. They are thought to represent a short phase after the WC phase.

Today we know that WR stars are the evolved descendants of massive O-type stars, with initial masses beyond 20$M_{\odot}$. In the evolution scheme of massive stars, first proposed by \citet{conti}, the two main WC and WN varieties of WR stars occur in evolutionary sequences as follows: when the initial mass $M>90M_{\odot}$  :  O $\rightarrow$ Of $\rightarrow$ LBV $\rightarrow$ WN $\rightarrow$ WC $\rightarrow$ SN (IIn). In lower mass ranges, e.g. 
25$M_{\odot}$<$M$<40$M_{\odot}$ the sequence is: O $\rightarrow$ RSG $\rightarrow$ WN $\rightarrow$ WC $\rightarrow$ SN (Ib).
In these examples (adopted from \citet{2003ARA&A..41...15M}), Of denote O supergiants with strong line emission and the initial stellar masses are only indicative, as they depend on metallicity.
More recent discussions of the evolution of WN and WC stars can be found in \citet{2012A&A...540A.144S} and \citet{2012ARA&A..50..107L}. 

WR stars themselves generally have masses between 10 - 25 $M_{\odot}$, and very high effective temperatures, typically from 25 000 K up to 100 000 K \citep{2007ARA&A..45..177C}. These stars spend approximately 10$\%$ of their 5 Myr lifetime in the WR phase \citep{2005A&A...429..581M}. Their characteristic, unusually broad emission lines are Doppler broadened due to the high velocities in their stellar winds, with terminal velocities $v_{\infty}$ typically between 1000 to 4000 km s$^{-1}$ (see \citet{2000A&A...360..227N} and references therein). These fast stellar WR winds are the strongest of all hot, luminous stars. Their mass-loss rates range around $\sim 10^{-5}$ $M_{\odot}$ yr$^{-1}$, an order of magnitude higher than massive O stars. This means that the feedback of these stars into the interstellar medium (ISM) is important for the chemical enrichment of their surroundings, especially since WR stars ultimately turn supernovae.
 
\citet{1972A&A....20..333A} attributed excess in infrared emission around some WC stars to thermal emission from dust grains at temperatures between 900 K and 1200 K. The existence of dust in these systems was puzzling because the high temperatures and luminosities (order 200000 $L_{\odot}$ for typical WC stars) should be efficient at destroying dust grains. A sublimation temperature of 2200 K is often assumed for amorphous carbon, and \citet{1972A&A....20..333A} found that with the observed luminosity, the central object must suffer six orders of visual extinction to allow dust survival at these temperatures. Furthermore, due to the lack of hydrogen in the wind of WC stars, grain formation is expected to be significantly less efficient around these stars \citep{2000A&A...357..572C}.
 
\begin{figure}
        \centering
       \includegraphics[width=\columnwidth]{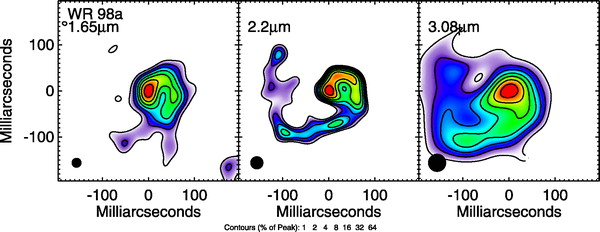}
        \caption{Keck I high angular resolution multiwavelength imaging of the pinwheel nebula WR98a, for three wavelengths as indicated, along with an indication of the beam width. Reproduced from~\citet{2007ApJ...655.1033M}, their Fig.~7.}
        \label{monnier}
\end{figure}

Meanwhile, many WR stars are suspected to host active locations of dust formation. These stars are typically late type WC stars. In 1995 a total of 26 were known, and \citet{1995IAUS..163..335W} introduced a classification dividing these stars in 19 ``persistent dust makers'' and 7 ``episodic'' dust makers. The \textit{Galactic Wolf-Rayet Catalogue}\footnote{\url{http://pacrowther.staff.shef.ac.uk/WRcat/index.php} (v1.14 accessed on 25 November 2015)} contains 63 WC stars with confirmed or suspected dust features out of a total of 634 WR stars. 
Nevertheless, more than forty years after the first observation of dust creation around WC stars by \citet{1972A&A....20..333A}, the physical conditions and the chemical reactions involved in forming dust around these stars are still unclear. Research on dust creation around single WC stars has been unable to provide clear answers: e.g. \citet{2000A&A...357..572C} find that the chemistry in the wind of WC stars is highly density dependent, and dust precursors are only created in significant amounts in very high density conditions (with gas number densities $n_{gas} > 10^{11}$ cm$^{-3}$). A viable, accepted model, at least for some well studied WR systems, is that of a binary system in which the companion star also has a stellar wind. In these cases, the wind collision region (WCR) can locally lead to strong compression, possibly enhanced by strong radiative cooling. In the WCR, material from a hydrogen-rich companion can be mixed with carbon-rich WC material. All of these factors would enhance dust formation. Evidence supporting this mechanism comes from the detection of strong X-ray and radio emission, suggesting the existence of a WCR in dust forming WR star WR 140 \citep{1990MNRAS.243..662W}. In this WR 140 system, as well as in a small number of other WR stars (e.g. WR 112 imaged with Keck by \cite{2007ApJ...655.1033M} or WR 118 as reported in~\cite{2009A&A...506L..49M}), infrared emission from dust is observed to form rotating ``pinwheel nebulae'', supporting that binarity plays a role in dust creation. The analysis of~\cite{2007ApJ...655.1033M} encompasses next to WR 104, WR 98a and WR 112, another 8 dusty Wolf-Rayet systems, to conclude on indirect evidence for linking dust shells in WC Wolf-Rayet stars to binarity. Figure~\ref{monnier}, taken from~\cite{2007ApJ...655.1033M} (their Figure~7), shows Keck images of WR 98a on UT 2000 June 24, at 1.65, 2.2, and 3.08 $\mu\mathrm{m}$, with the beam spot in the lower left corner indicating the resolution loss towards longer wavelenghts.

Dust formation is thus currently believed to happen in the wake of the wind-wind collision cones, at distances a few hundred stellar radii from the hot evolved massive binaries, as already reviewed in~\cite{vanderHucht2001}. Early models investigating dust formation in hot stellar winds for WC-type Wolf-Rayet stars pointed instead to a dense, neutral, circumstellar disk~\citep{Cherchneff1995}, to circumvent the seeming impossibility to form dust in a spherically symmetric, low-density, ionized, hot outflow. Recent observations of supernova SN2008ax with AKARI/IRC detected reradiation from dust that formed around the OB/WR progenitor in an interacting binary system~\citep{Sakon2010}. The same work also reports on near-infrared imaging and spectroscopic observations of WR 140 (made in 2009), directly observing ongoing dust formation in these systems.
In this sense, there is no longer any doubt that dust can form under the harsh conditions encountered in wind-wind collision zones. In this paper, we will adopt this well accepted model, and investigate differences in the gas dynamic behavior, depending on how effective local radiative losses can be inside these colliding wind scenarios.

To enhance our understanding about the physical properties of the environment around WR stars, we here set forth to model a colliding wind binary involving a WR star on scales from the near-stellar outflow up to roughly a thousand AU. We aim to investigate the formation and structure of the WCR, as well as how the large scale dust formation and redistribution happens in these binaries. To do so, we will use the Wolf-Rayet binary WR 98a to guide our modelling efforts.

\subsection{The Wolf-Rayet 98a system}

WR 98a, also referred to as IRAS 17380-3031, was suggested as a WR candidate by \citet{1989AJ.....98..931V}. This was confirmed by \citet{1991ApJ...378..302C} who identified it as a late type WR star. It was already clear that the star has significant circumstellar dust emission. In \citet{1995MNRAS.275..889W} the classification is further narrowed to either the WC8 or WC9 subclass. \citet{1995MNRAS.275..889W} also derive a wind velocity of $\sim 900$ km s$^{-1}$ from the width of the 1.083 $\mu$m He \textrm{I} 2p-2s emission line. They remark that in photometric observations spanning between March 1990 and July 1992, the emission due to dust varies by about half a magnitude in all near-infrared bandwidths observed (J, H, K, L and M). They conclude that the dust production around WR 98a may be variable but is clearly persistent. From variability in photometry they deduce a periodicity of $\sim$ 1.4 yr. Later, \citet{1999ApJ...525L..97M} used aperture-masking interferometry at the Keck I telescope to make multi-epoch high spatial resolution ($\lesssim$50 mas) images of WR 98a at 2.2 $\mu$m. Their results revealed that WR 98a forms a ``pinwheel nebula'' (see figure~1 in \cite{1999ApJ...525L..97M}), the second such discovered, following the first episodic dust maker WR 104 \citep{1999Natur.398..487T} (see also later observations of WR 104 in \citet{2008ApJ...675..698T}). The pattern for WR 98a was explained as the result of an interaction with a binary companion, probably an OB star. Assuming typical values for a WC9+OB binary with the mentioned periodicity results in a binary separation $a \sim 4$ AU. An Archimedean spiral pattern with constant angular velocity (corresponding to a circular orbit) fits the observations at all epochs, from which \citet{1999ApJ...525L..97M} derive that the orbit is not highly eccentric. Interestingly, the Archimedean spiral fit yields an estimate for the orbital period of 565$\pm$50 days (1.55 yr), in good agreement with the periodicity of 1.4 yr found from the variation in photometry. This variation can be due to a small eccentricity, or may also be due to the viewing angle (the spiral is inclined 35$^\circ\pm 6^\circ$ to the line of sight). We will show evidence in favor of this viewing angle interpretation. A (near) circular orbit can be expected if one of the stars was significantly larger in the past (the binary separation is in the same order of magnitude as the radius of a late-type red supergiant) and tidal interactions or a common envelope phase resulted in orbital circularisation. The determination of the period is further refined by \citet{2003IAUS..212..115W}, who find a periodicity of 564 $\pm$ 8 days from photometric observations.

\begin{figure}
	\centering
	\includegraphics[width=\columnwidth]{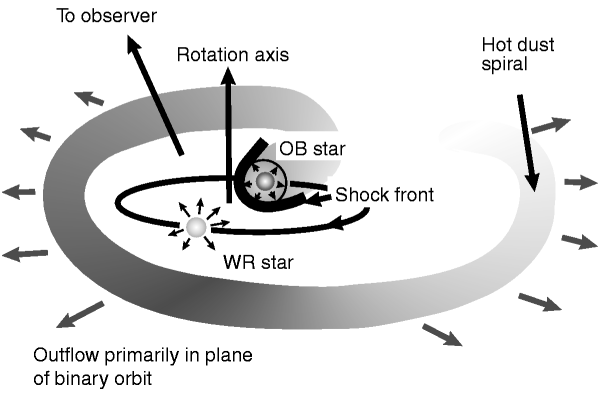}
	\caption{Cartoon view on a binary WR+OB system, such as the WR 98a system. Reproduced from~\citet{1999Natur.398..487T}.}
	\label{cartoon}
\end{figure}

\citet{2002ApJ...566..399M} studied WR 98a in the radio domain using the Very Large Array (VLA). They find evidence of nonthermal emission, an expected sign of wind-wind collisions. From the inferred thermal emission, an estimate of the mass-loss rate is made, giving $\dot{M} = 0.5\times10^{-5}$ $M_\odot$ yr$^{-1}$.
As will be quantified further on, a WC9 related wind loss dynamically dominates over the outflow of the OB companion star, which means that the expected wind-wind interaction pattern due to the orbital motion is one of a spiral pinwheel marking the outflow zone of the OB star, as schematically indicated in Fig.~\ref{cartoon}. Our numerical simulations will demonstrate that the actual interaction indeed persists to very large scales, but the detailed morphology of the interaction zone turns out to be heavily dependent on the effectiveness of radiative cooling.

\subsection{Previous binary wind modeling}


Before we give the details on our model for studying the dynamics and dust redistribution in WR 98a, we give a short overview of previous numerical investigations of colliding-wind binaries and their conclusions. In so doing, we especially emphasize length and timescales covered, as they typically constitute a major difference with our approach.

\citet{1992ApJ...386..265S} used 2D simulations to study the wind collision region in binaries of early-type stars. Cooling is included in these simulations, and the effect of several instabilities on the collision region is discussed in detail. The orbital motion is ignored in these simulations, and only the central region close to the stars is modelled. The authors model two observed systems, including the WC7 + O4-O5 binary WR 140 (at periastron). The X-ray luminosity and spectra of the collision region is also modelled.

Walder and Folini performed several important early (3D) investigations of the hydrodynamics of colliding wind binaries \citep{1999IAUS..193..298W,2000Ap&SS.274..189F,2003IAUS..212..139W} and the stability of colliding flows in (radiative) shocks \citep{1995IAUS..163..525W,2000Ap&SS.274..343W}. 

\citet{2004MNRAS.350..565H} modelled the dust-producing binary WR 104 by applying a radiative-transfer code to a spiral density profile created from a simple ballistic particle model. They find that their model compared well with the SED, finding a dust production rate of 8$\pm1 \times 10^{-7}$ $M_\odot$ yr$^{-1}$, or 2.7 $\%$ of the WR mass-loss. With a size of only 10 nm the dust grains used are relatively small, however they note that their adopted model is relatively insensitive to grain size and they cannot exclude the necessity of larger grains.

In \citet{2009MNRAS.396.1743P} the 3D hydrodynamics of O+O binaries is investigated, including radiative cooling as well as the acceleration of the wind itself. The grid in the simulation is half-cubic (reflected over the $xy$ plane), with sides $x$ and $y$ both spanning 7.5$a$ in the largest model, with $a$ the semi-major axis of the orbit. A resolution of 456$^3$ is used in the largest model. Amongst other results, slow moving cold clumps are found to exist for a substantial amount of time in the WCR when the post-shock is largely adiabatic.

\citet{2011A&A...530A.119P} modelled WR22, a WN7+O binary, in 3D using AMR (5 levels, effective resolution 4096$\times$4096$\times$2048). The simulated domain has a size of 9.5$a\times$9.5$a\times$4.8$a$. Their investigation focussed on the dynamics of the inner WCR and X-ray emission in the high-temperature shock. To do so they take radiative cooling into account, and made an investigation of the importance of including actual wind acceleration and radiative wind breaking. They find that including wind acceleration and breaking reduces the stability of the WCR in WR22 significantly.

\citet{2011A&A...527A...3V} used 3D hydro with inclusion of radiative cooling to study two binary systems, one being a hydrogen-rich WR star (subtype WNL) + O star and the other a LBV + O star. The comparison underlines the importance of including radiative cooling. Due to the difference in physical properties the shock morphology is radically different for the two binaries, with the former being smooth and free of instabilities while the radiative shock in the latter is prone to thin-shell instabilities \citep{1983ApJ...274..152V,1994ApJ...428..186V}. The used grid has a size of $4a\times4a\times0.4a$ using four levels of mesh refinement to achieve an effective resolution of $480\times480\times80$ cells.

In \citet{2012A&A...546A..60L} an extensive parameter study is performed to investigate the role of the momentum flux ratio of the winds $\eta = (\dot{M}_{1} v_{1})/(\dot{M}_{2} v_{2})$ and the velocity difference on the large scale stability of the spiral structure. The simulations for the parameter study are performed in 2D. No radiative cooling is included. They find that when the velocity difference in the winds is strong, depending on $\eta$ the configuration can become unstable (in 2D), due to Kelvin-Helmholtz (KH) instability. A case study based on the dust producing binary WR 104 is also made, in which the 2D simulations are compared to a smaller scale 3D simulation (effective resolution 4096$^3$ using six AMR levels on a (12$a$)$^3$ box, simulated for 1/8 of an orbit).

More recently, \citet{2015A&A...577A..89B} performed a 3D simulation of the interaction between the winds of an O star and a pulsar for one orbital period on scales beyond the binary separation (domain size $64a\times 64a\times 20a$). The pulsar wind is chosen to be mildly relativistic (Lorentz factor $\Gamma=2$). Many similarities to non-relativistic colliding wind binaries are seen, however the extreme velocity difference causes strong instabilities: of (mixed) Kelvin-Helmholtz, Rayleigh-Taylor (RT), and Richtmyer-Meshkov (RM) type, leading to turbulence formation at large distances. They note that the onset of instabilities is much stronger in 3D than in 2D, which enhances shocks and material mixing. Instead of using AMR a nested grid is used (the inner $4a \times 4a \times 2a$ region has a fixed $256 \times 256 \times 128$ resolution, and the outer region has a fixed $512 \times 512 \times 256$ resolution). Also, the cells in the outer region are mapped to an exponentially stretched grid. The equivalent resolution of the grid is 4096$\times$4096$\times$2560.

Complementary to all these efforts, we here present a model where both the (shock-governed) gas dynamics and the dust insertion and redistribution, is studied in detail. Our 3D gas plus dust evolution is further fed into an advanced radiative transfer code, allowing us to produce synthetic images at infrared wavebands, or even mimic `photometric' observations. We first provide all details on the numerical setup and how it connects with the WR 98a system.

\section{Physical and Numerical setup}
\subsection{WR 98a parameters}\label{physSetupwr98a}

From the observations of WR 98a we build a physical model as follows. We adopt a WR star of a WC9 subtype and use a mass of $M_{WR} = 10 M_\odot$ \citep{2012A&A...540A.144S}, mass-loss rate $\dot{M}_{WR} = 0.5\times10^{-5}$ $M_\odot$ yr$^{-1}$ \citep{2002ApJ...566..399M}, and wind velocity $v_{WR,\infty} = 900$ km s$^{-1}$ \citep{1995MNRAS.275..889W}. The OB type companion is less constrained by the observations, but we assume $M_{OB} = 18 M_\odot$, $\dot{M}_{OB} = 0.5\times10^{-7}$ $M_\odot$ yr$^{-1}$, and $v_{OB,\infty} = 2000$ km s$^{-1}$, consistent with recent atmosphere models for massive OB stars \citep{2015PASP..127..428F}. This fixes an important dimensionless parameter in the dynamics of colliding wind binaries, namely the momentum flux ratio $\eta$ of the winds
\begin{equation}
\eta = \frac{\dot{M}_{OB} v_{OB,\infty}}{\dot{M}_{WR} v_{WR,\infty}} = 2.22\times10^{-2}. \label{momBalance}
\end{equation}
This low value of $\eta$ indicates that the outflow of the WR star dominates. Using the stellar masses and the orbital period of $P_{\mathrm{orb}}=565$ days \citep{1995MNRAS.275..889W,1999ApJ...525L..97M,2003IAUS..212..115W}, the semi-major axis length $a$ of the orbit is calculated to be 
\begin{equation}
a = \sqrt[3]{\frac{P_{\mathrm{orb}}^2 G (M_{WR} + M_{OB})}{4 \pi^2 }}= \, 6.08\times10^{13} \, \text{cm} \simeq 4.06 \,\text{AU}, \label{semimajor}
\end{equation}
with $G$ the gravitational constant. As observations suggest that the orbit is (nearly) circular \citep{1999ApJ...525L..97M}, we set the eccentricity of the orbit to zero. The temperature for the winds in both stars is set to $10^4$ K, justified from assuming that photoionisation keeps the unshocked winds at these temperatures. The terminal wind radii, being the distance to the stars where their wind has been accelerated to the fixed terminal velocity, is set to $r_{WR,\infty} = 5\times10^{12}$ cm and $r_{OB,\infty}= 3\times10^{12}$ cm for the WR and OB star, respectively. Initially, both stars are embedded in a static, gas-only medium with a density $\rho_{\mathrm{ISM}} = 10^{-20}$ g\,cm$^{-3}$ and a temperature of $T_{\mathrm{ISM}}=10^4$ K, similarly photoionized by surrounding stars. 
The wind zones of both stars are introduced in the domain using two moving internal boundary regions, which are at all times spheres of radii $r_{WR,\infty}$ and $r_{OB,\infty}$ within which the stellar wind reaches its terminal velocity. In practice, cells inside these regions are always replaced by the theoretical density, pressure and velocity profile, i.e. for the WR star a density profile with $\rho(r_{WR})=(r_{WR,\infty}/r_{WR})^2\dot{M}_{WR}/(4\pi r_{WR,\infty}^2v_{WR,\infty})$ (where $r_{WR}$ denotes the instantaneous radial distance to the centre of the WR star), a constant purely radial velocity of magnitude $v_{WR,\infty}$, at the fixed temperature of $10^4$ K. For reference, the densities for both winds at their terminal radii are of order $\rho_{WR,\infty}={\cal{O}}(10^{-14})$ g\,cm$^{-3}$ and $\rho_{OB,\infty}={\cal{O}}(10^{-16})$ g\,cm$^{-3}$. Note further that the terminal speeds are highly supersonic: Mach numbers are 76 and 170 for the WR and OB star, respectively.
 
The instantaneous positions of the stars (and thus of the internal boundaries) have the stars follow their exact Keplerian orbit, which are circular in the $xy$-plane. Initially, both stars are along the $x$-axis, and we solve Kepler's equation such that the motion of the stars is counter-clockwise in all simulations. 

In our numerical approach, we rescale all physical values to code units, for which we choose as reference for time $t_0$ the orbital period $P_{\mathrm{orb}}=4.89\times10^{7}$ s (1.55 yr), for length $L_0$ the semi-major axis $a=6.08\times 10^{13}$ cm, and a density $\rho_0 = 10^{-16}$ g\,cm$^{-3}$. All other scalings can be derived from these, e.g. the reference mass is $\rho_0 L_0^3$, a reference speed $v_0$ is $L_0/t_0$, while a reference temperature value is given by $T_0= k_B/(m_H v_0^2)$ (using Boltzmann constant $k_B$ and proton mass $m_H$).

\subsection{Governing equations}\label{WRnumSetup}  

We use the gas+dust hydrodynamics module of the {\tt MPI-AMRVAC} code \citep{2012JCoPh.231..718K,2014ApJS..214....4P} for modelling the dynamics around the WR binary on an adaptive mesh. This module allows to treat in a fully self-consistent manner the (shock-governed) evolution of gas and (multiple) dust fluids which interact with the gas by means of (dust grain size dependent) drag forces. The method used for representing a size distribution by multiple pressureless dust fluids is benchmarked in \citet{2014ApJS..214....4P} and demonstrated in a molecular cloud context in \citet{Hendrix2015}. Earlier circumstellar models using this approach already indicated that important differences in the dynamics of grains of different sizes can be found, with smaller dust grains effectively coupled to the gas, while larger grains can even surpass the bow shock associated with a stellar wind bubble~\citep{vanMarle2011}. In the present work, the computational challenge to cover the ${\cal{O}}(1000)$ AU scales for multiple orbits of the binary system is considerable, and we therefore incorporate only one dust fluid, necessarily omitting such details.
For the very same reason, we can not afford to perform full resolution studies on 3D domains as large as we cover here. We did perform parametric surveys, in both physical as well as resolution space, for 2D runs, representative for conditions in the equatorial plane of our current model. In~\cite{Hendrix2015WRproc}, we show that by adjusting the momentum flux ratio parameter according to $\eta_{2D}=\sqrt{\eta_{3D}}$, we can forcibly obtain the same stand-off position for the wind-wind collision region in 2D mockup runs. These 2D runs have helped to pinpoint the requirements on resolution for the 3D runs discussed in this paper. At the same time, we refer for actual convergence studies on a dusty gas test suite, partly inspired by those introduced for smoothed particle hydrodynamical simulations of gas-dust mixtures by~\cite{Laibe2012} to section~3 of~\cite{2014ApJS..214....4P}.

In practice, the following equations are solved 
\begin{eqnarray}
\frac{ \partial \rho}{\partial t}  \quad + & \nabla \cdot (\rho \mathbf{v}) & 
    = S^{\mathrm{int}}_{\rho,WR-OB} - S^{\mathrm{mix}}_{\rho},\label{continuityEquation}\\
\frac{ \partial (\rho \mathbf{v})}{\partial t}  \quad + & \nabla \cdot (\rho \mathbf{v} \mathbf{v}) + \nabla p & =  \mathbf{f}_d +\mathbf{S}^{\mathrm{int}}_{\rho\mathbf{v},WR-OB} -\mathbf{S}^{\mathrm{mix}}_{\rho\mathbf{v}}, \label{momentum} \\
\frac{ \partial e }{\partial t}  \quad + & \nabla \cdot \left[ (p+e) \mathbf{v} \right] & =  \mathbf{v}  \cdot\mathbf{f}_d - \frac{\rho^2}{m_H^2} \Lambda (T)  \nonumber\\
&& \,\,\, +S^{\mathrm{int}}_{e,WR-OB} -S^{\mathrm{mix}}_{e}, \label{energy}  \\
\frac{ \partial \rho_d}{\partial t}  \quad + & \nabla \cdot (\rho_d \mathbf{v_d}) & = S^{\mathrm{mix}}_{\rho_d}, \label{dust}\\
\frac{ \partial (\rho_d \mathbf{v_d})}{\partial t}  \quad + & \nabla \cdot (\rho_d \mathbf{v_d} \mathbf{v_d}) 
& =  - \mathbf{f}_d +\mathbf{S}^{\mathrm{mix}}_{\rho_d \mathbf{v_d}}, \label{momentumdust} \\
\frac{ \partial \theta_{WR}}{\partial t}  \quad + & \mathbf{v} \cdot \nabla \theta_{WR} & = S^{\mathrm{int}}_{\theta_{WR}} , \label{tr1} \\
\frac{ \partial \theta_{OR}}{\partial t}  \quad + & \mathbf{v} \cdot \nabla \theta_{OR} & = S^{\mathrm{int}}_{\theta_{OR}} . \label{tr2} 
\end{eqnarray}
In these equations, $\rho$, $\mathbf{v}$, $p$, and $e$ are the mass density, velocity, pressure, and total energy density of the gas, where the $\gamma$ = 5/3 constant polytropic index enters the relation $e  = p/(\gamma - 1) + \rho v^2/2$.  
Furthermore, $\rho_d$ and $\mathbf{v_d}$ denote the mass density and the velocity of the dust fluid. The final two equations (which are actually solved in an equivalent conservative form) introduce two tracer quantities that identify the WR and OB wind zones. These are used to define the mixing zone, where dust is created, as parametrized by the source terms indicated with $S^{\mathrm{mix}}$.
All source terms indicated with the {\em{int}} superscript like $S^{\mathrm{int}}$ denote sources that enter solely through our use of an internal boundary for both wind zones, as explained above. 
The term with a temperature dependent cooling curve $\Lambda(T)$ on the right hand side of equation~(\ref{energy}) represents the loss of energy due to optically thin radiative losses. In our models, we will contrast simulations with these losses omitted versus runs where they are incorporated. In the outflow of WR stars, the high metallicity strongly enhances the radiative cooling. We use an adequate table by \citet{2002A&A...394..901M} (as seen in their figure 1). The implementation of the cooling algorithm is explained in detail in \citet{2011CF.....42...44V}, we use the exact integration method as introduced by~\cite{Townsend}. The force $\mathbf{f}_d$ represents the drag force of the gas on the dust fluid with density $\rho_d$. This dust fluid is treated as a pressureless fluid coupled to the gas by the drag force.
The drag force we adopt is the Epstein drag \citet{1924PhRv...23..710E,2012A&A...545A.134M}, 
\begin{equation}
\mathbf{f}_d = (1-\alpha(T)) \sqrt{\frac{8 \gamma p}{\pi \rho}} \frac{\rho \rho_d}{\rho_{p} a_d} ( \mathbf{v_d} - \mathbf{v} ), \label{dragforce}
\end{equation}
with $\rho_{p}$ the grain material density of the dust fluid and $a_d$ the grain size. The sticking coefficient $\alpha$ is a temperature dependent factor given by
\begin{equation}
	\alpha = 0.35 \exp{\left(-\sqrt{\frac{TT_0}{500 \mathrm{K}}}\right)}+0.1.\label{sticking}
\end{equation}
This sticking coefficient is a measure for the efficiency of the momentum transport in the collisions between dust grains and gas, and is adopted from \citet{2006A&A...456..549D}. Note that our pressureless treatment has no dust temperature, so only the gas temperature $TT_0$ enters here (the multiplication with $T_0$ restores the dimensions in this evaluation). In expression~(\ref{sticking}) we make the assumption that the dust is (much) cooler than the gas, which is the case in the work presented here. This will also be verified a posteriori, when turning our simulations into virtual infrared images using a radiative transfer code, which will quantify a temperature acquired by the dust through illumination from the central stars.

\subsection{Numerical setup}
The coupled gas+dust fluid equations and both tracer equations are advanced using a total variation diminishing Lax-Friedrich (TVDLF) scheme with the strong stability preserving Runge-Kutta SSPRK(5,4) 
five-step, fourth order time discretisation, which is stable up to CFL of $\sim$1.508 (\citet{SpiteriRuuth2002}, see \citet{2014ApJS..214....4P} for the implementation in {\tt MPI-AMRVAC}). Nevertheless, we do limit the timestep by using a CFL number of 0.3 for gas and dust, and also evaluate a separate dust acceleration criterion based on the stopping time of dust grains \citep{Laibe2012}. The selected spatio-temporal discretization of the eleven equations is of a conservative finite volume type, with all quantities interpreted as cell-center values representing cell averages of the conservative quantities. In evaluating the fluxes across cell edges, a limited linear reconstruction is used to achieve a higher (spatially second) order in smooth regions. Due to the extreme length scales covered, we actually always employ primitive variable limiting, where furthermore densities and pressures are transformed to a logarithmic scale before their reconstruction. The limiter is of a monotonised central (MC) type \citep{1977JCoPh..23..263V} for the first 10 refinement levels, while on the finest mesh level ($l_{max}=11$), a minmod limiter is used instead.

The physical domain is taken as [$x$,$y$,$z$] $\in$ [-160$a$,160$a$;-160$a$,160$a$;-70$a$,70$a$], with $a$ the semi-major axis length of the orbit. For WR 98a, this means a box of about 1300 AU on the sides, and 570 AU vertically. The resolution is chosen in a way that the small-scale structures in the WCR close to the stars are resolved, while computing the large scale structure of the WCR such as the spiral and the possible onset of instabilities. To do so, the use of adaptive mesh refinement (AMR) is crucial: we use an effective resolution of 81920$\times$81920$\times$24576 by using eleven AMR levels (80$\times$80$\times$24 on the coarsest level). In so doing, our finest grid cell is of size $0.0039a\times 0.0039a \times 0.00569a$, resolving structures down to ${\cal{O}}(10^{11})$ cm. Through the use of a finite volume discretization, we have the freedom to make the cells non-cubic, and the use of 11 AMR levels means in practice that cell volumes can differ by a factor of $8^{10}$ (but the proper nesting of the AMR levels makes sure that the grid hierarchy is such that neighboring blocks/cells at most differ by one level, i.e. a factor 8 in volume). This also means that our internal boundary zones are covered by at least 20 cells through the sphere diameter (in fact, the WR wind zone is covered by 2.6$\times 10^4$ cells). The outer boundaries are treated as open boundaries, using a Neumann boundary with vanishing derivatives on all quantities.

Due to the extreme effective resolution, we need a suitable refine as well as derefinement strategy. The refinement is done using a L\"ohner estimate (i.e. a weighted evaluation of the instantaneous discretized second derivative, as in \cite{lohner}) using (1) the logarithm of the density, and (2) the local tracer product $\theta_{WR}\theta_{OB}$ in a 30-70 \% weighing scheme. We further ensure the highest level mesh is always present in each wind zone, extended up to 1.3 times the wind terminal radius. A gradual trend to coarsen to a maximum of 9 levels is based on geometric arguments (beyond $0.25a$ from each star). The most important derefinement is in line with our aspiration to study the inner WCR in high resolution, and keeps the total number of refined blocks under control by gradually limiting the maximal refinement level with increasing distance to the binary system. For all coarse grid levels $l\in[1,6]$, we quantify a distance $D(l)$ given by
\begin{equation}
D(l) = 155.0-153.5 \log \left[ -9.0\left(\frac{l-1}{l_{max}-4}\right)^2+10 \right] \,.
\end{equation}
From this distance $D(l)$ on, we only allow at most level $l_{max}-2-l$ to be activated. In practice, this realizes a gradual derefinement from levels 8 down to 3 within $42.5\,a$ from the binary. All regions further than $42.5\,a$ from the binary are always coarsened to levels of at most the third level. However, for the run with radiative cooling, we actually enforced refinement up to level 3 everywhere, for all times when dust insertion took place (hence levels 1 and 2 are unpopulated for the last 3 orbital periods). 

\subsection{Fluid tracers and dust formation}
\label{dustFormWR98a}

The addition of two tracer fluids $\theta_{WR}$ and $\theta_{OB}$ advected with the flow allow to track the material ejected by the WR star and the OB star, respectively. The values of the tracers are set to 100 in the internal boundary regions. At $t=0$, the tracers are zero everywhere else. At all later times, the local tracer values range between $0$ and $100$, as each tracer distribution will have a (numerically diffused) boundary demarking the star's wind zone. Since the same discretization is used for the dynamical quantities in our model, we can use the local tracer product $\theta_{WR}\theta_{OB}$ as an indication of where effective mixing of material from the WR and OB stars has occured (see also \citet{2012A&A...546A..60L}). The existence of such a mixing layer is of high importance for dust formation, as it enriches the WR material with hydrogen. When the product reaches 2500, we can loosely identify this with a region filled with 50\% WR material and 50\% OB material. This type of identification is used in our semi-empirical model for dust formation as follows.

We set forth to reproduce the eventual outcome of dust nucleation and growth processes as close as possible to what is known from observations. Since our goal is to study the combined large and small-scale dynamics (at the WCR), affecting the dust redistribution, we parametrize the source terms $S^{\mathrm{mix}}$ by matching known quantities such as the dust formation rate, the grain size (distribution), and the dust formation location. The latter aspect incorporates that (1) dust forms in locations where the dense, H-poor WR wind is enriched with H-rich material from the OB outflow, and (2) that the dust production rate is larger in regions with higher densities. 
For the grain size distribution, our (single) dust fluid represents the same grain size distribution as found for WR 118 by \citet{2001A&A...379..229Y}, where the number density of grains of size $s$ has a dependency $n(s) \propto s^{-3}$ with $s$ between 5 nm and 600 nm. The use of a single bin to represent this distribution translates in a dust grain radius $a_d=15.44 \times 10^{-7}$ cm. It is to be noted that this choice for the representative dust grain radius is obtained when one wants to ensure that the drag force correctly captures the total force on the adopted grain size distribution between 5 and 600 nm: this means that this drag force (and its size dependence) enters as a weighting function, and then sets the size of the representative dust grain, a procedure which we clearly explained in section 3.2 of~\cite{Hendrix2014}. In both the dynamical and post-process radiative transfer simulations, these dust grains are modelled as amorphous carbon grains as found by observations \citep{1987A&A...182...91W,2002ApJ...565L..59M}, and thus they have a material density $\rho_p = 2$ g cm$^{-3}$ \citep{2002ApJ...565L..59M}. Using our dimensionalization as stated earlier, this makes the factor $\rho_p a_d= 5.09 \times 10^{-4}$ in the drag force formula~(\ref{dragforce}).

To introduce dust in the WR outflow we combine the global information about dust growth from observations with information our hydrodynamical simulations provide. Our simulations start from a pure gas dynamics setup and steadily introduce dust in certain locations. It is reasonable to assume that dust is most significantly produced in the WCR and in the outflow of the WCR because (1) high density clumps are formed there through compression and possibly further density increases due to thermal instabilities, and (2) because in this region hydrogen from the companion is mixed in the dense (H deprived) WR wind, which significantly boosts dust nucleation. To add dust in the WCR, we need to know the grain size (distribution) and grain material density, but also the allowed region of dust formation. The latter we restrict geometrically to be between distances $D_{in}$ and $D_{out}$ from the binary. 
Observations allow us to put some constraint on these aspects. In \citet{1987A&A...182...91W} the inner location of dust formation is modelled for several WR stars, using the assumption of a spherically distributed dust shell. The fitted inner shell radii for WC9 stars has an average value of $\sim$340 R$_{*}$ (R$_{*}$ being the stellar radius), which corresponds to $\sim$10.5 AU using R$_{*}$ = 6.6 $R_{\odot}$ for WC9 stars. For WR 140, a highly eccentric ($e=0.84$, P=7.94 years; \citet{1990MNRAS.243..662W})  WC7 + O4-5 binary, observations reveal that dust only is formed close to the periastron passage. WR 140 has been considered the prototype for episodic dust formation around WC stars. \citet{1990MNRAS.243..662W} derive that dust is only formed at 2400 stellar radii ($\sim$150 AU) from the central objects. For our WR 98a model, we will adopt an inner radius of $D_{in} = 20.3$ AU, which is a reasonable value given these observational facts.
Finally, we need the local dust formation rate $\phi$ which will enter the source terms $S^{\mathrm{mix}}$. As we explain in an appendix, the value $\phi=0.0763$, together with values $D_{in} = 20.3$ AU and $D_{out} = 212.8$ AU can be shown to translate in a realistic dust production rate of $10^{-8}$ $M_\odot$ yr$^{-1}$, which is 0.2 \% of the total mass loss rate of WR 98a \citep{2002ApJ...566..399M}. These three semi-empirical parameters $\phi$, $D_{in}$ and $D_{out}$ enter our dynamical model ensuring dust formation at a correct rate. In simplified ``pseudocode'', the model quantifying the source terms $S^{\mathrm{mix}}$ addition within time step $\Delta t$ can be written in the following way: 

\begin{lstlisting}[mathescape]
if (time > duststart) then
  $\Delta \rho_d = \rho \phi \Delta t$
  FormDust = False
  
  if ($D_{in}$ < distance < $D_{out}$) then
    if (($\rho_d+\Delta \rho_d$) > minimal dust density) then  
      if (in mixing layer) then
         FormDust = True
      end if
    end if
  end if      
  
  if ( FormDust == True ) then
    $\rho_d^{new} = \rho_d^{old} + \Delta \rho_d$
    $\rho^{new} = \rho^{old} - \Delta \rho_d$  	 ! conserve total mass
    $v^{new}_{d}=v^{old}_{gas}$             ! inherit gas velocity
    $e^{new} = e^{old} - \Delta \rho_d  (v^{new}_d)^2 / 2$ ! keep pressure fixed 
    $[\rho_d v_{d}]^{new} = [\rho_d v_{d}]^{old} +  \Delta \rho_d v^{new}_{d}$
    $[\rho v_{gas}]^{new} = [\rho v_{gas}]^{old} -  \Delta \rho_d v^{new}_{d}$ 
  end if
  
end if   
\end{lstlisting}  
This pseudocode emphasizes that dust insertion will start only after a finite time, which we take two or three orbital revolutions. Moreover, we only create dust in locations identified by the geometric and dynamic (mixing layer) criteria. The latter is defined by $\theta_{WR}\theta_{OB}>500$ as explained in the appendix. If a floor value for detecting dust presence is exceeded (this is set to $10^{-11}$ in dimensionless units), we extract gas matter and turn it into dust, conserving total mass and momentum. It also maintains the velocity of the gas and the pressure of the gas to avoid artificial triggering of gas dynamical instabilities. 

As explained in appendix, our hydro plus gas model handles the macro-physics, where it is quite appropriate to consider a situation where the (majority) of the gas that is not converted to dust behaves essentially adiabatic. Indeed, our model cannot take into account small scale clumping (as stated before, the smallest cell width in our model is order $10^{11}$ cm), which may in fact be instrumental in creating the conditions necessary for dust formation (optically thick, cold gas), even when the gas as a whole still behaves adiabatically.

\section{Results}

\subsection{Hydrodynamics}

\begin{figure}
        \vspace{-0.0cm}
        \centering
        \includegraphics[width=0.8\columnwidth]{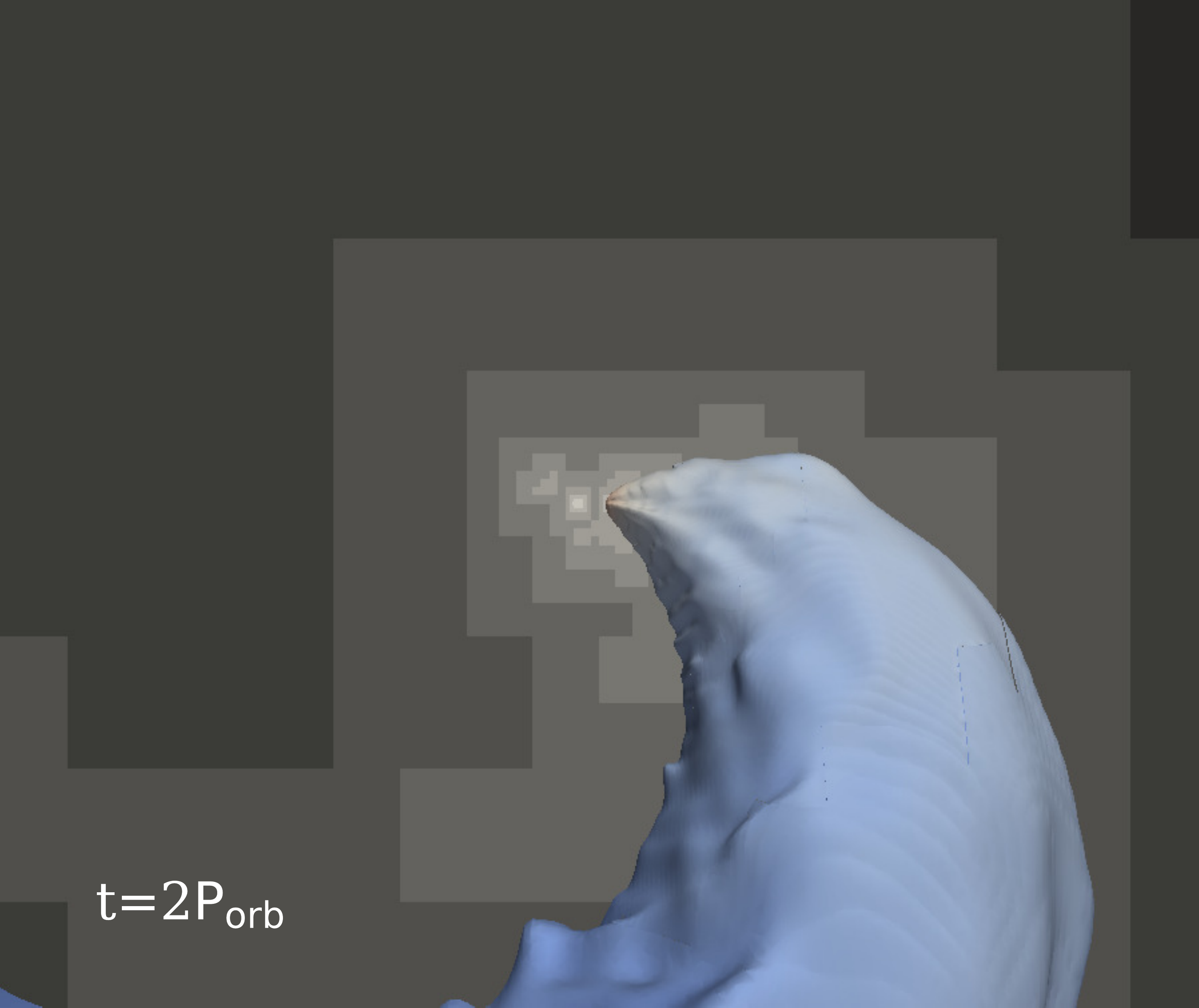}
        \includegraphics[width=\columnwidth]{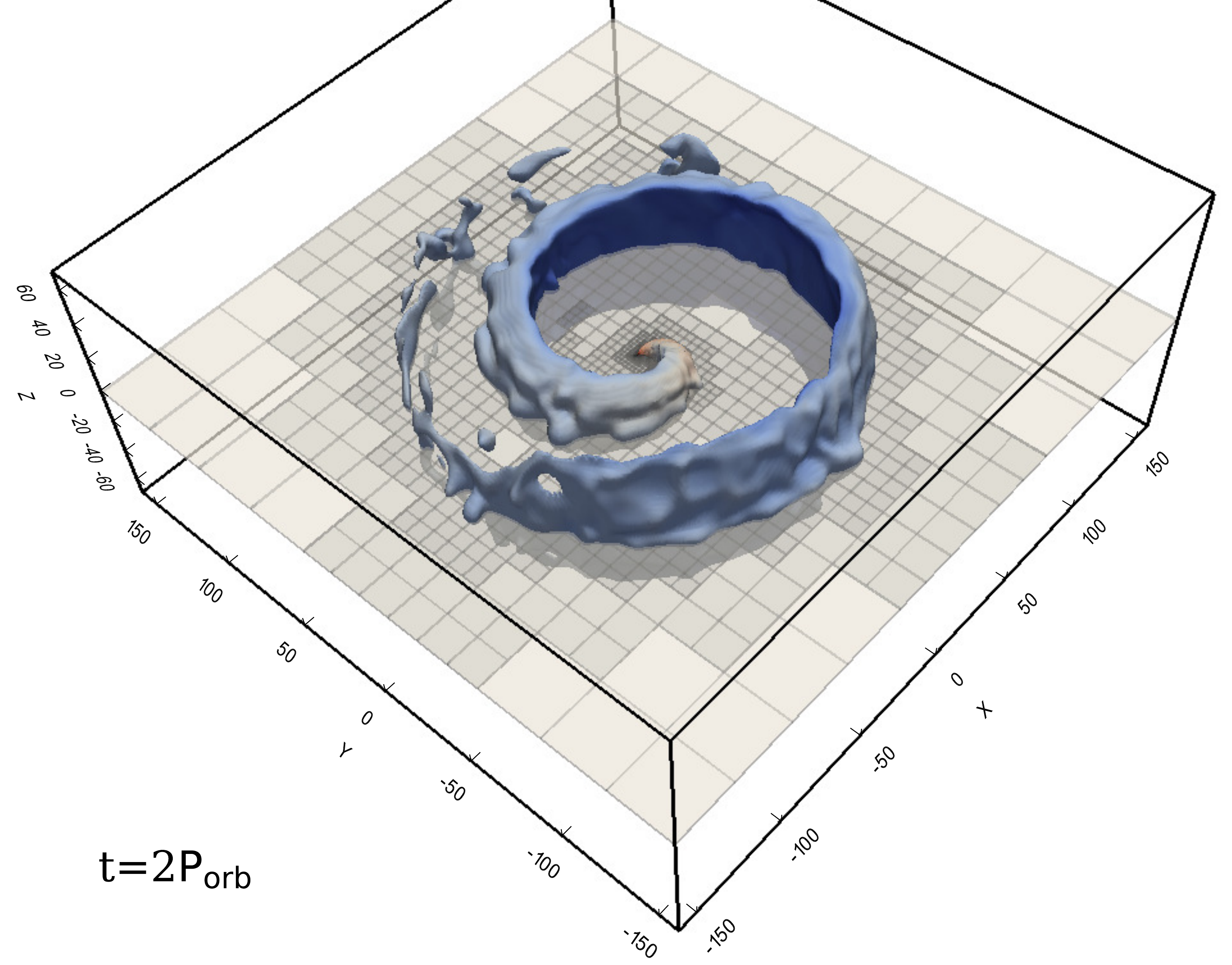}
        \includegraphics[width=\columnwidth]{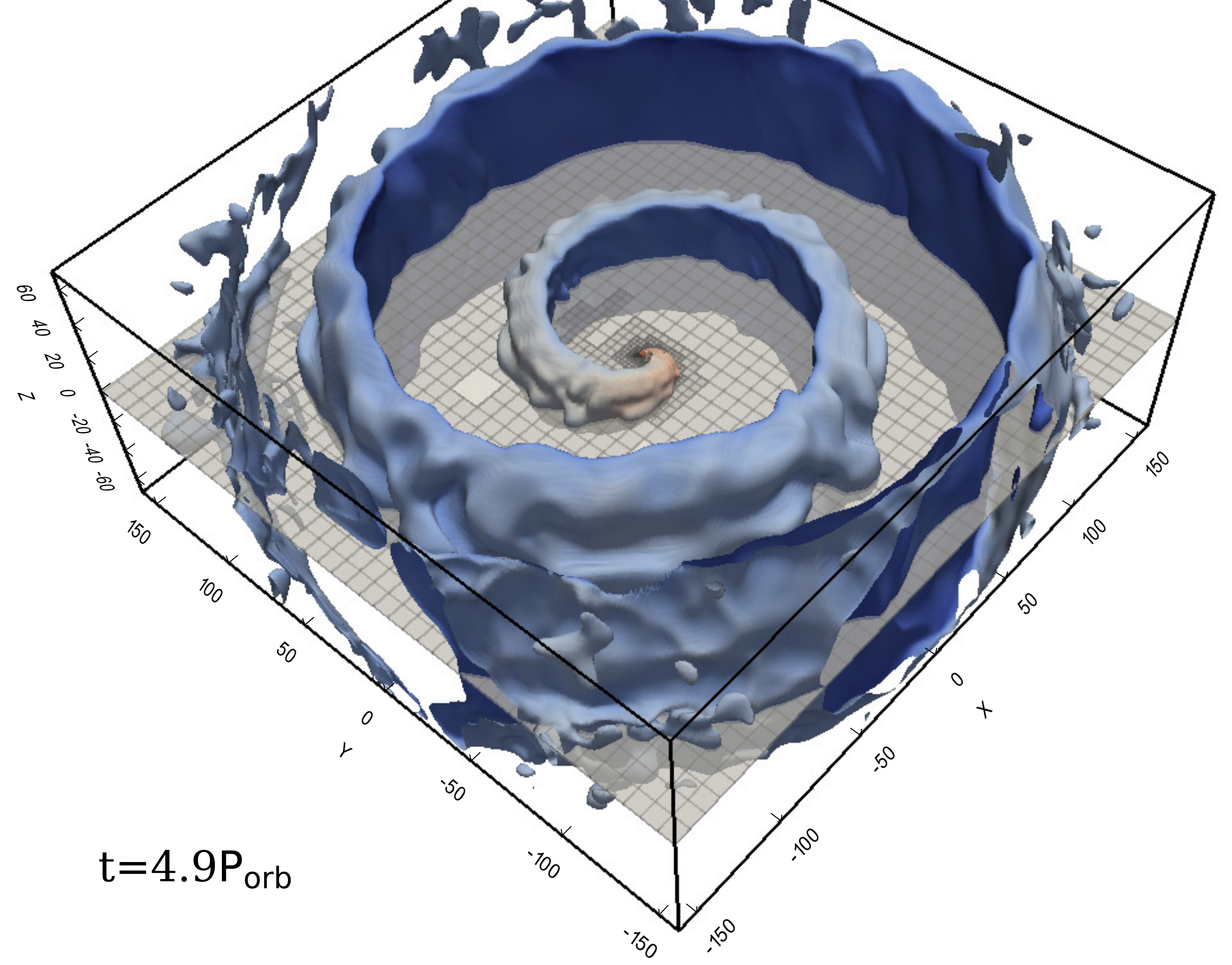}
        \caption{The large-scale structure of a binary WC9+OB star system like WR 98a. We render the adaptive grid structure and the mixing layer (with a density colored isosurface of the mixing degree) at two (3.1 years) and after 4.9 (or 7.6 years) orbital periods of the binary. From the $t=2 P_{\mathrm{orb}}$ timeframe onwards, dust creation is allowed for the following three orbits. In middle and bottom panel, the entire simulation domain is shown, with distances measured in multiples of the semi-major axis $a$ ($\sim$ 4 AU). An animated version is provided as well.
Top: A zoomed view on the central region shows the highest grid levels only, for $t=2$.} \label{gridstuff}
\end{figure}

In this section, we describe the overall hydrodynamical evolution of the circumbinary medium. We will contrast an adiabatic case where no radiative cooling is acting, with one where cooling is incorporated. We note that there is a slight inconsistency in our use of the term `adiabatic', since both scenarios have a (small) fraction of the gas converted to dust in a scenario as described earlier in this paper, meaning that there is a sink of mass and energy, also in the `adiabatic' case. We will however stick to this nomenclature further on, as it qualifies the gas evolution quite appropriately. The first few orbital periods (two without cooling, three with cooling) are simulated without any dust insertion, and merely serve to establish the dynamic wind interaction zone over a sufficiently large part of the domain. In the case without cooling, Fig.~\ref{gridstuff} demonstrates that one gets the expected cartoon view from Fig.~\ref{cartoon}, since the region of significant mixing forms a clear spiral pattern in the wake of the OB star. Shown in Fig.~\ref{gridstuff} is the configuration obtained after two orbits without dust insertion (top panels), as well as the final spiral pattern found at the end of our run (4.9 orbital periods). The top panel is a zoomed top-down view on the central region, emphasizing the extreme scale-resolving capacity of the AMR run: the finest 7 levels concentrate in the wind zones and the WCR, while the full domain view shows how the outer region is still employing level 1 grid blocks all along the outer borders. The spiral-shaped isosurface identifies the mixing layer ($\theta_{WR}\theta_{OB} = 500$), which together with the geometrical constraints ($D_{in}$ and $D_{out}$) controls the dust creation. Clearly, after two orbits, the mixing zone is easily identified over the entire region where dust will be formed, i.e. up to the distance $D_{out}\approx 52.5 a$. By this time, the average speed attained in the mixing zone, and also throughout the region invaded by both wind zones, has increased and stabilized at about 850-900 km/s, consistent with the dominance of the WR outflow. The mixing layer isosurface in Fig.~\ref{gridstuff} is colored by the local density distribution (in an arbitrary logarithmic scale), and it is seen to decrease further along the spiral pattern. The endstate at 4.9 orbits resolves the mixing layer throughout the entire 1300 AU lateral dimensions: several of the windings have by then left the open boundaries without notable artificial reflections.

\begin{figure}
        \centering
       \includegraphics[width=\columnwidth]{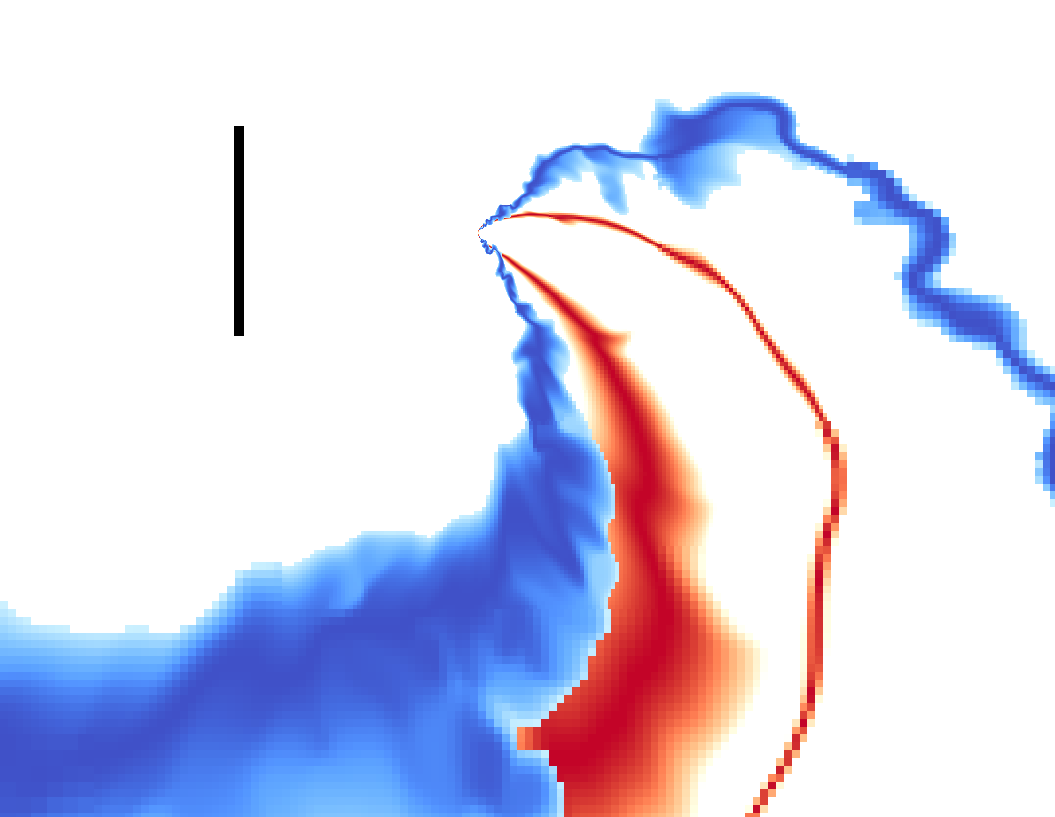}
        \caption{Slice through the orbital plane of two 3D simulations at $t=2$. The simulation with cooling, which has a larger opening angle, is shown in blue, while the simulation without cooling is shown in red. For both simulations we show where significant mixing has taken place ($\theta_{WR}\theta_{OB} > 500$). Only the central part of the simulation is shown, the black ruler has a length of 10$a$.}
        \label{3dMix}
\end{figure}

\begin{figure}
        \centering
       \includegraphics[width=\columnwidth]{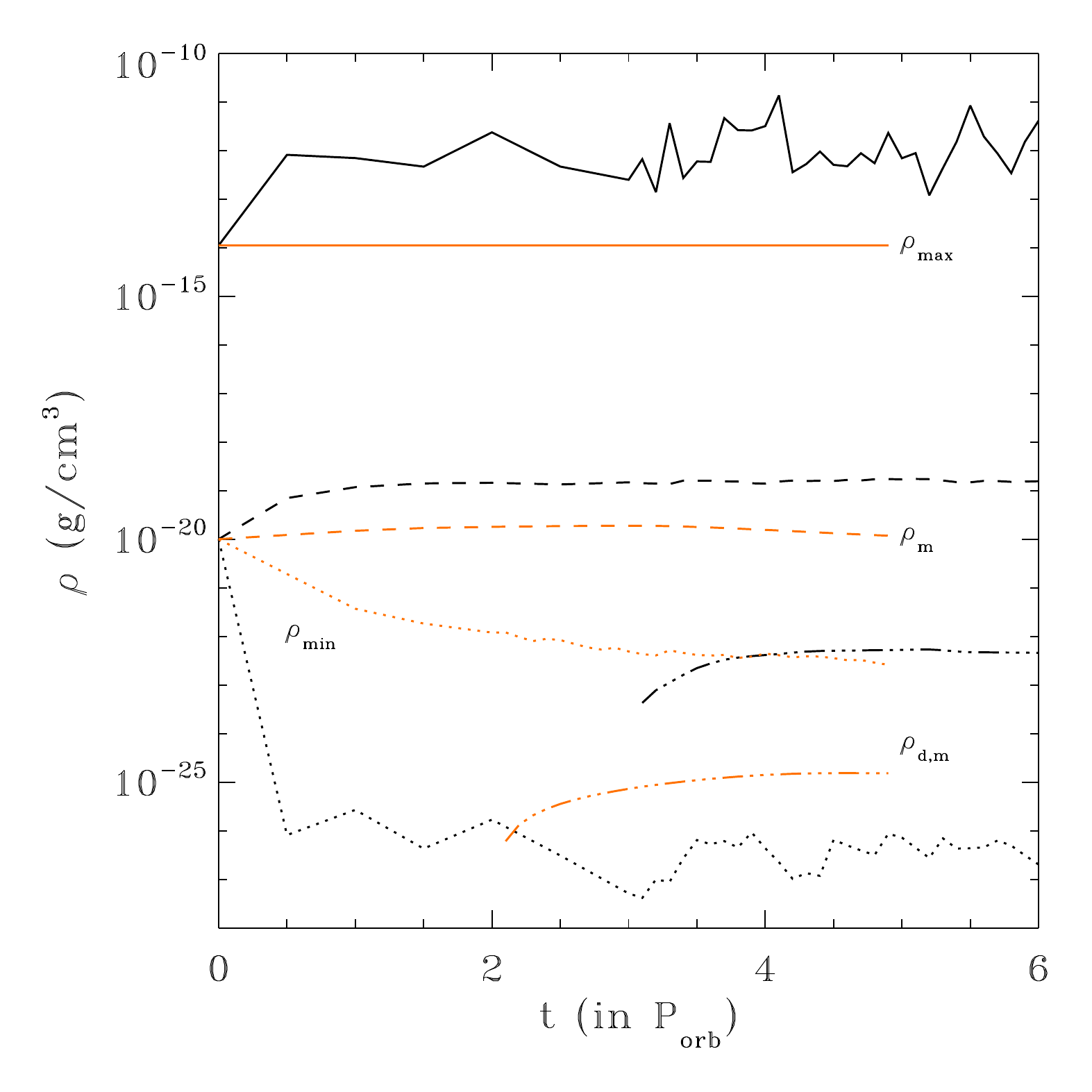}
        \caption{Evolution of the extremal gas densities (maxima as solid lines, minima as dotted lines) for both simulations as function of orbital period. The dashed lines give the mean gas density $\rho_m$ as function of time, the dash-triple-dot line quantifies the mean dust density $\rho_{d,m}$ obtained in the simulation. The latter starts after 2 and 3 orbital periods, for the case without and with cooling, respectively.}
        \label{figcomp}
\end{figure}

In Fig.~\ref{3dMix}, we compare the inner WCR structure for a simulation with cooling (blue) and a simulation without cooling (red). Both are shown after two orbital periods, and the shaded areas indicate the regions of significant mixing $\theta_{WR}\theta_{OB} > 500$ in the orbital plane. A first observation is that both cases demonstrate a clear asymmetry between the leading and the trailing flank of the mixing zone: the leading flank is narrower than the trailing zone, and is also an area of higher prevailing gas densities. A second observation is that the case with cooling (blue) clearly shows a more turbulent structure. As will be discussed further on, thin shell and thermal instabilities dominate the hydrodynamics in the WCR when cooling is allowed. These enhance the mixing and result in the turbulent motion of the WCR visible in Fig.~\ref{3dMix}. Interestingly, our simulations show that cooling also causes a change in the opening angle of the WCR. An approximation (accuracy $\sim 1\%$) of the semi-opening angle $\theta$ (in radians) is given by \citet{1993ApJ...402..271E}:
\begin{equation}
\theta = 2.1 \left( 1 - \frac{\eta^{2/5}}{4} \right) \eta^{1/3} \qquad \text{for } 10^{-4} \leq \eta \leq 1.
\end{equation}
Since for our setup $\eta = 2.22\times10^{-2}$ as in equation~(\ref{momBalance}), we would expect a semi-opening angle of $\theta \sim 32^{\circ}$. This is indeed in accord with the roughly $60^\circ$ arc seen in the central region. In the simulation with cooling, we clearly see a larger semi-opening angle of approximately $\theta=45^{\circ}$, further down the spiral. The dynamic WCR zone when cooling is active thereby effectively enlarges the volume over which mixing occurs. This, together with the enhanced density contrasts resulting from the thermal instability developments, will make the case with cooling a much more efficient dust creator. This is confirmed by Fig.~\ref{figcomp}, where the temporal evolution of the extremal gas density values, along with the mean average density for both gas and dust is given for both cases (with and without cooling). The extremal densities shown are those found on the entire domain, excluding the (prescribed) wind zones of both stars. Therefore, for the case without cooling, we find the maximal gas density at all times (shown by solid lines) to correspond to the terminal wind density of the WR star, of order $10^{-14}$ g cm$^{-3}$. However, we see that for the case with cooling, the instantaneous maximal density can go up by two to three orders of magnitude, and is highly variable. The instantaneous minimal gas density is shown as a dotted line for each case, and extremely low density values (below $10^{-26}$ g\, cm$^{-3}$) are encountered in the case with cooling, while the density mimima in the case without cooling gradually decrease from the initial $10^{-20}$ g\, cm$^{-3}$ down to
$10^{-23}$ g\, cm$^{-3}$. Overall, the mean gas density in the entire volume $\rho_m=\int\rho \,dV/V$ (shown with dashed lines) stays close to $10^{-20}$ g\, cm$^{-3}$ without cooling, while it is up by one order of magnitude in the case with cooling. The extreme gas density contrasts encountered in the case with cooling are the result of especially thermal instabilities causing large density differences. The mean dust density $\rho_{d,m}$ is shown for both cases as a dash-dotted line as well (only starting from $t=2$ without cooling, and from $t=3$ with cooling). Clearly, more gas is being converted to dust in the case with cooling included.

\begin{figure}
        \centering
        \includegraphics[width=\columnwidth]{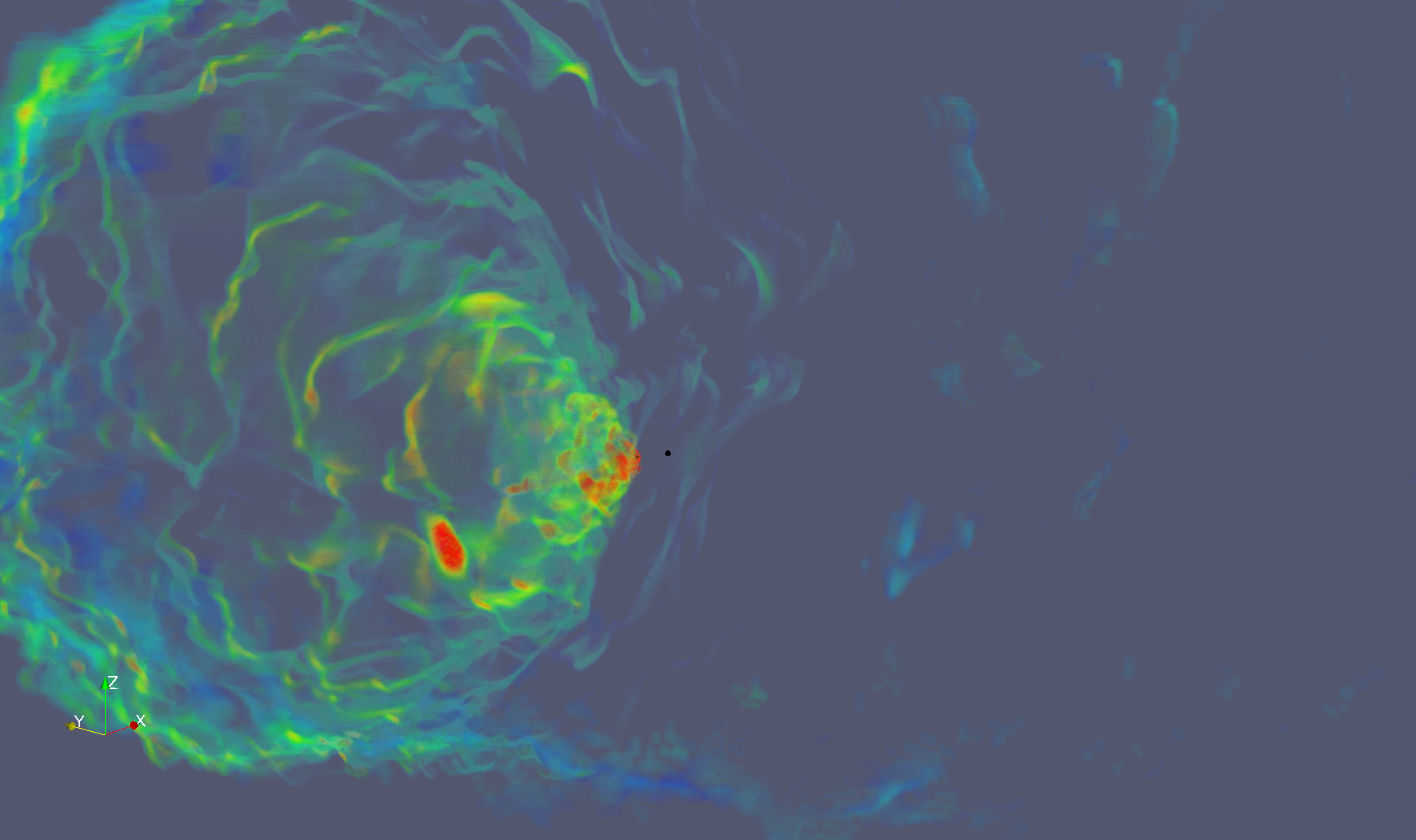}
        \caption{3D volume rendering of the gas density structure in the WCR in a simulation with cooling at $t=0.43$. Only the central part of the domain is shown. The terminal wind radii of both stars are marked in black, however the O star is obscured by high density (red and green coloured) clumps in this figure. The terminal wind radius of the WR star is visible in the centre of the image. It contains $\sim 2.5\times10^4$ cells. An animated version of this figure is provided.}
        \label{3dWRClumps}
\end{figure}

\begin{figure}
        \vspace{-0.0cm}
        \centering
        \includegraphics[width=\columnwidth]{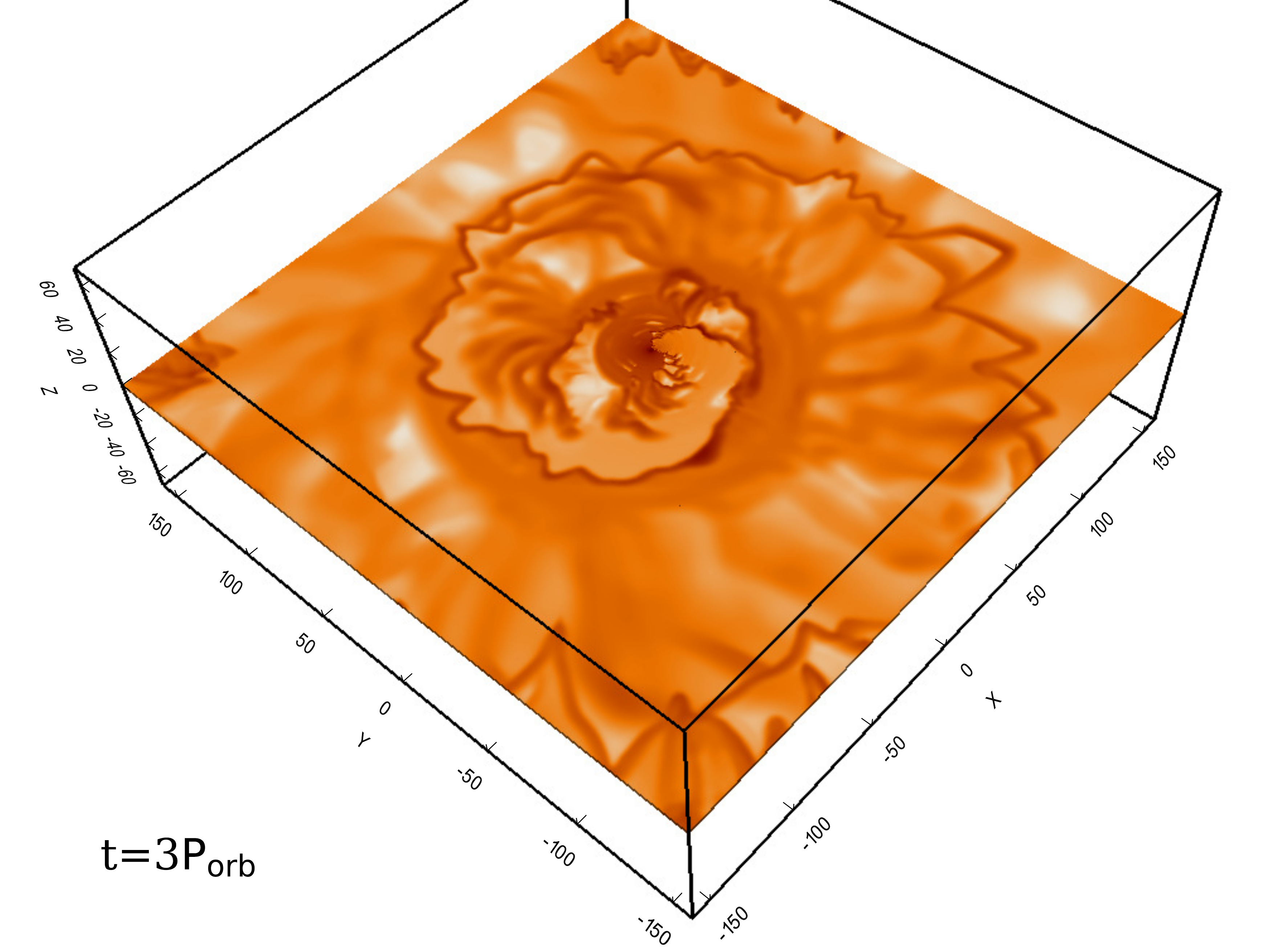}
        \includegraphics[width=\columnwidth]{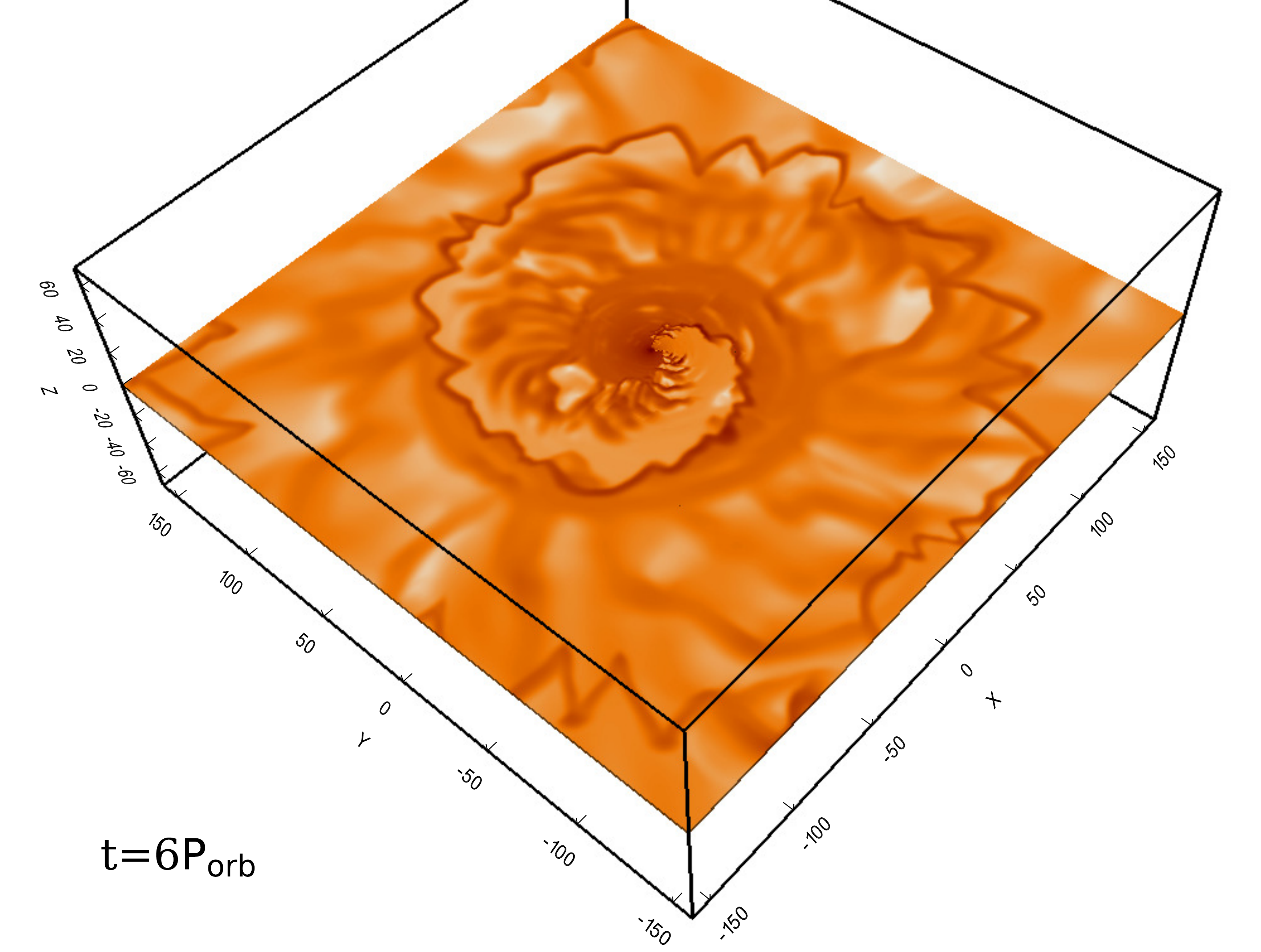}
        \caption{Top: A slice showing the density structure in the orbital plane at $t=3$ for the simulation with cooling. From this point onwards, dust creation is allowed for the following 3 orbits. Bottom: idem, but after 6 orbits.}\label{coolcase}
\end{figure}

Figure~\ref{3dWRClumps} shows a 3D visualisation of the gas density structures in the WCR for the case with cooling. High density clumps and filaments formed by thermal instability are clearly visible (an animation of this figure is provided as well). The color scale in Fig.~\ref{3dWRClumps} is adjusted to properly visualize the extreme density range discussed earlier. The gas density structure is in fact dominated by localized clumps of high density material (red) and instability development is occuring all around the WCR. In fact, when we quantify the instantaneous volume of where dust is allowed to form, i.e. the region between the concentric spheres defined by $D_{in}<r_{WR}<D_{out}$ and occupied by highly mixed material (i.e. having $\theta_{OB}\theta_{WR}>500$), we find a fairly constant total domain volume fraction of about 0.023. Thus, about half of the full 0.042 volume fraction defined by the spheres alone is liable to dust production. In contrast, the case without cooling has the same volume of dust formation roughly constant at 0.004, stating that only 10 \% of the full concentric spherical region is then containing highly mixed matter (roughly where the spiral pattern intersects the spheres). The mixing of the two winds is thus much more effective with cooling. This is also seen in the density structure in the orbital plane, shown at times $t=3$ and $t=6$ in Fig.~\ref{coolcase}. The highly dynamic WCR still shows an overall winding pattern, but large zig-zag patterns are superimposed, as the thermal instabilities induce pressure variations, leading to even more turbulent motions. This is also reflected in the overall expansion velocities of the mixed wind zones: in the case with cooling the entire mixing area has an average velocity of up to 1400 km/s, while the average temperature in the dust formation zone ultimately decreases below $10^5$ K.
The case with cooling shows little evidence for shear-flow related instabilities, but the density views from Fig.~\ref{3dWRClumps}-\ref{coolcase} confirm that effective radiative cooling influences the structure of the WCR. The strong cooling appropriate for the high metallicity in the WR outflow induces a very turbulent morphology of the WCR, such that the local densities can be orders of magnitude denser than in adiabatic evolutions. Finally, the mixing and compression of material from both stars over a larger region causes significantly more material to be present in the mixing layer and thus liable to dust creation. We conclude that radiative cooling introduces significant changes to the physical environment of the WCR. These differences are especially important for dust formation, which would be enhanced due to the presence of more hydrogen enriched material in dense clumps. The clumps may also shield the dust grains from the hard radiation of the massive stars. The dust morphology will also be very clumpy (following the gas density contrasts to some extent), although we still can discern a very disturbed `spiral' shape in the density views of Fig.~\ref{coolcase}.

The situation is rather different for the adiabatic case.
We stated earlier that for that case without cooling, the average speed in the mixing zone stabilized at about 850-900 km/s. In fact, when we quantify the average speed in the dust formation zone alone (hence $D_{in}<r_{WR}<D_{out}$ and $\theta_{OB}\theta_{WR}>500$) for this adiabatic case, we find 1200 km/s. This is consistent with the fact that high velocity, low density material is injected by the OB star. This dust formation region for the adiabatic case is also compressionally heated to a high $1.5\times 10^6$ K temperature. 
All these factors make the case without cooling one where a clear spiral mixing zone develops around the contact discontinuity between both winds. On the leading edge (the binary motion is anti-clockwise in the simulation) the mixing stays closely bound to the contact discontinuity, while on the trailing edge a broad mixing layer forms. In the adiabatic (no cooling) case, the velocity difference between the two wind regions causes also velocity shearing near the contact discontinuity and the region is potentially prone to Kelvin-Helmholtz instabilities. 
After about one spiral winding, the mixing layer along the trailing contact begins to show evidence of the onset of Kelvin-Helmholtz instability. Nevertheless, when evaluated on larger scale, the spiral pattern stays stable up to large distances. This is the reason why the leading edge of the spiral is indeed showing more fine-structure, although the overall spiral structure expands without getting fully destroyed. While the motion of the wind material is mostly radial, the transverse velocity ($\vec{v_{t}} = \vec{v} - (\vec{v} \cdot \hat{r}) \hat{r}$), is almost parallel to the spiral pattern at larger distances. While the velocity of the OB material is intrinsically higher due to the higher terminal velocity, it is interesting to note that the confined OB material has a stronger transverse component: the OB wind zone has deviations from radial speeds where the fraction $v_t/v$ is of order 2 to 3 \% on the inner windings, going up to 10 \% on the outermost winding found within our simulation domain. The WR zone has a velocity vector almost entirely radial.

By using the method described in section \ref{dustFormWR98a} to quantify the appropriate dust formation rate, we introduce dust in the 3D simulations after few orbits, when the stellar outflow has pushed away sufficient material and a representative part of the outflow has formed all over the dust formation zone. We can see from Fig.~\ref{gridstuff} (adiabatic) and Fig.~\ref{coolcase} (with cooling), where the time of dust activation and the final state are compared, that the introduction of dust does not dramatically alter the overall outflow morphology. When we further investigate where the dust accumulates, the case with cooling shows rather pronounced fragmentation due to the high density contrasts. We will return to this case when discussing virtual radio images. For the remainder of this section, we concentrate on the adiabatic case showing the clear spiral pattern (i.e., related to Fig.~\ref{gridstuff}).

\begin{figure}
        \vspace{-0.0cm}
        \centering
        \includegraphics[width=\columnwidth]{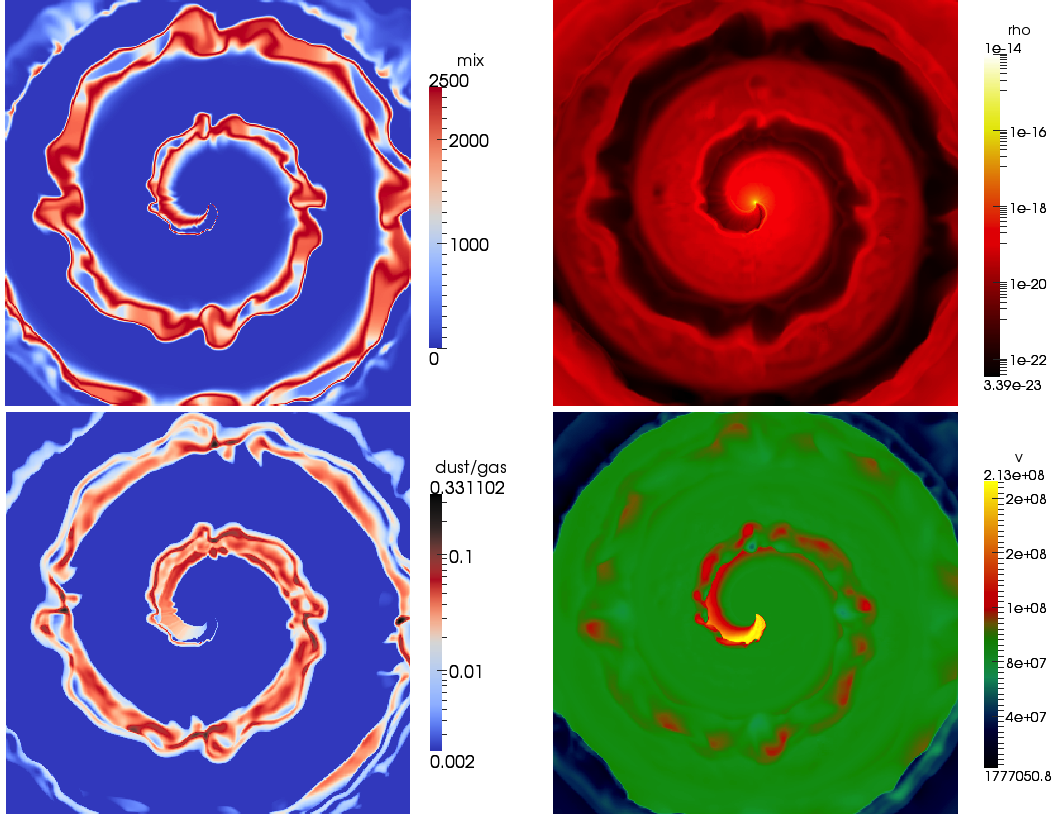}
        \caption{Four slices through the orbital plane in a 3D simulation with dust and without cooling, at $t=4.9$. Slices perpendicular to the plane are shown in Fig.~\ref{2SliceY=0AtT=4_9_v2}. The four panels show the wind mixing (top left), gas density (top right), dust-to-gas density ratio (bottom left), and total gas velocity (bottom right). The full $xy$ domain of the simulation ($\left|x\right|<160a$ and $\left|y\right|<160a$) is shown.}
        \label{4facetsAtT=4_9}
\end{figure} 

\begin{figure}
        \vspace{-0.0cm}
        \centering
        \includegraphics[width=\columnwidth]{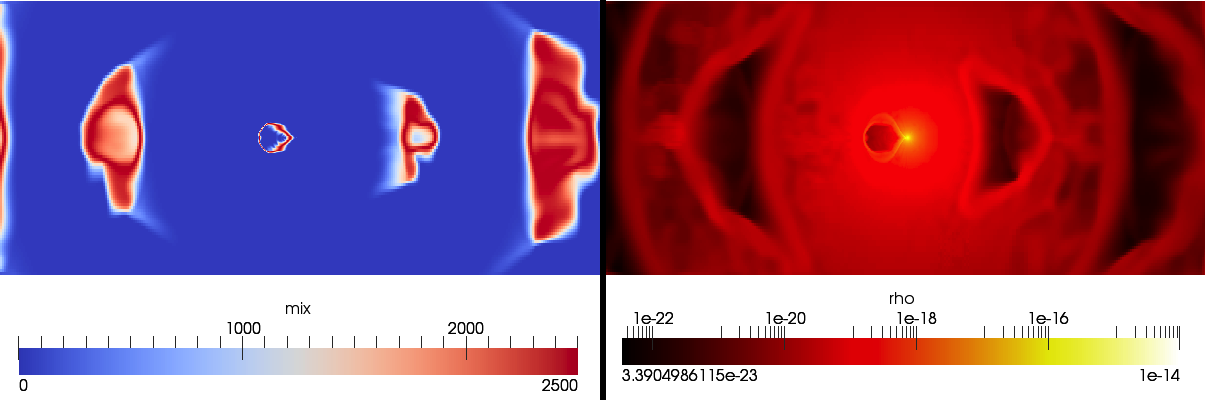}
        \caption{Two slices at the same time and of the same simulation as in Fig.~\ref{4facetsAtT=4_9}, but sliced along the $x=0$ plane. The entire $yz$ domain of the simulation is shown. The left panel indicates the mixing of material from the two winds. The right panel shows the density distribution of the gas.}
        \label{2SliceY=0AtT=4_9_v2}
\end{figure}

Figure \ref{4facetsAtT=4_9} shows the outcome of the adiabatic simulation at $t=4.9$, or after about 3 orbits with dust creation activated. Shown in this figure are several slices quantifying the variations in the orbital plane. Figure \ref{2SliceY=0AtT=4_9_v2} shows two slices made at the same time as Fig.~\ref{4facetsAtT=4_9}, but perpendicular to the orbital plane. The 3D simulation displays some bending in the spiral pattern as seen in the orbital plane, and this is related to the shear flow effects mentioned earlier: especially the leading edge of the spiral can show clear deviations from a smooth spiral pattern. Bending of this leading edge leads to compression of the OB star material, and at several locations the trailing and leading mixing layers even merge. As can be seen in Fig.~\ref{4facetsAtT=4_9} in the top left panel quantifying the mixing, the fast OB wind is being mixed with slower WR material, which reduces the velocity shear. This mixing is clearly visible when we use the wind tracers (top left panel), as well as by the deceleration it causes (bottom right panel, quantifying the total gas velocity). However, these effects are less evident in the density distribution itself (shown in the top right of Fig.~\ref{4facetsAtT=4_9}). In both Fig.~\ref{4facetsAtT=4_9} and Fig.~\ref{2SliceY=0AtT=4_9_v2} we see that the material of the OB star forms a spiral of underdense material in which mixing takes place. The mixing region is surrounded by a region of slightly enhanced density, most pronouncedly along the leading edge of the spiral. The bottom left panel of Fig.~\ref{4facetsAtT=4_9} shows the dust-to-gas ratio in the simulation. A closer inspection reveals several interesting features in the image. We see that dust starts to form close to the binary, from where it gradually increases the dust-to-gas ratio. The presence of dust can be seen to be closely linked to the regions where mixing has taken place. In most regions the stopping time for the dust is short and thus the distance that dust can travel independently from the gas remains small, and we therefore see that the dust is efficiently coupled to the gas in which it forms. Furthermore, the simulations reveal that in the entire spiral the dust-to-gas ratio is typically around $\sim 0.05$, with regions where the contrast goes up to $\sim 0.33$. These values are higher than those encountered in typical ISM conditions. In fact, they are high enough to cause potential dynamic effects on evolving Kelvin-Helmholtz instabilities, which our earlier work~\citep{Hendrix2014} has investigated in more idealized local conditions when several dust species are accounted for (but here we only have one dust species taken along).

\begin{figure}
        \vspace{-0.0cm}
        \centering
        \includegraphics[width=\columnwidth]{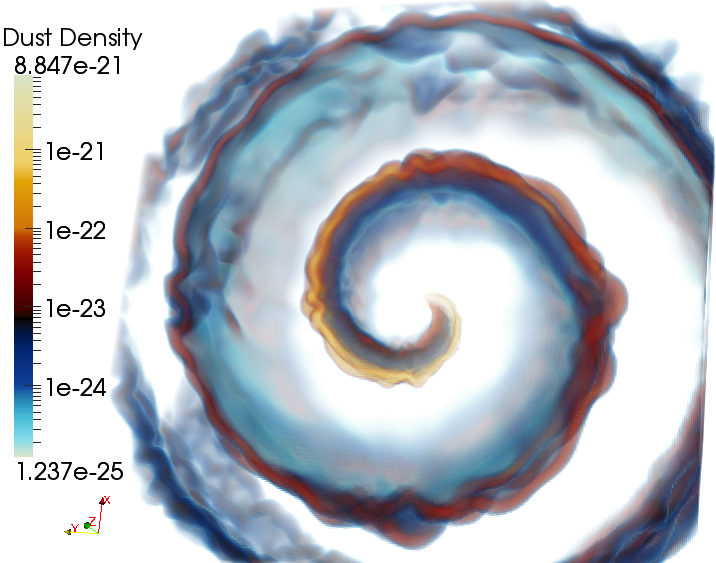}
        \caption{Volume rendering of the dust density distribution at $t=4.9$ in a simulation without cooling. The entire 3D domain of the simulation is shown.} \label{WR3DVolRender}
\end{figure}

The final dust density distribution itself is shown in a volume rendered fashion, for the adiabatic case at its endtime $t=4.9$ in Fig.~\ref{WR3DVolRender}. This view confirms that indeed dust has been distributed preferentially on the leading edge of the mixing layer that defines a spiral pattern. The trailing edge also contains dust, but in a more diffused morphology. There is clear fine-structure related to the (shear-flow related) undulations of the leading spiral. One can note that actually, the entire OB star wake is filled with dust material. Over the 1300 AU shown in the figure, dust-free zones remain clearly visible as well, with denser matter there that relates to the WR outflow zone. In the following section, we will discuss how this self-consistently computed dust distribution that results from wind-wind interaction in our OB+WR binary system will appear in actual infrared wavelength views.

\subsection{Synthetic observation}

\subsubsection{Radiative transfer using SKIRT}
Our hydrodynamical models covered three orbital revolutions for a dust-producing WR+OB binary environment, and obtained the resulting dust distribution on a scale of about 1300 AU in the orbital plane. This evolving dust distribution can then be observed synthetically. To this end, we post-process the instantaneous 3D dust distribution $\rho_d(\mathbf{x},t)$ with a Monte Carlo radiative transfer code SKIRT~\citep{2011ApJS..196...22B,Camps201520}. This open-source code simulates continuum radiation transfer in dusty astrophysical systems, and has e.g. been applied to study (virtual) galaxies (like M51 in \citet{2014A&A...571A..69D}), or to mimic conditions in specific molecular cloud regions~\citep{Hendrix2015}. The SKIRT code uses Monte Carlo recipes to emulate how photon packages ultimately get detected by a virtual telescope, after being emitted from a specified source (or sources), while on-the-way undergoing physical processes including scattering, absorption and (re-)emission by dust particles. The native octree AMR data structure from MPI-AMRVAC is one of many available input data structures recognized by SKIRT, and could be used for the radiative transfer calculation directly. 
However, the AMR grid refinement strategy employs the gas and tracer product densities, while the dust component is the relevant medium for our radiative transfer purposes. We therefore let SKIRT construct a new octree grid, recursively subdividing the spatial domain until the dust density distribution is properly resolved. For a typical run (e.g. for the final snapshot of the run with cooling), the MPI-AMRVAC grid has about 18 million cells organized in a block-adaptive fashion using 11 AMR levels, and the SKIRT octree dust grid has about 8 million cells (leaf nodes), 90 percent of which have an optical depth below 0.02. To make the virtual observations more realistic, the single representative grain in the MPI-AMRVAC dust density distribution $\rho_d(\mathbf{x},t)$ is at the same time converted to a mixture of 25 representative grains with different sizes, according to the same overall number density distribution as was used in MPI-AMRVAC (i.e. $n(s)\propto s^{-3}$ with grain size $s\in[5,600]$ nm). 
Note that in~\cite{vanMarle2011}, we pointed out that for the circumstellar environment of Betelgeuse, the larger grains (there from 5 to 250 nm) could be progressively decoupled from the location of the overarching shock front. That work used 10 dust species in a similar setup as the one adopted here, and indeed showed subtle differences in the morphology of dust grains, large to small. However, for practical reasons, we could not afford doing multiple dust bins in the hydro plus dust simulation of this paper. For the grain sizes reported here, we may expect similar differences, although the situation is rather different as we are considering a colliding wind zone (not a bow shock about a single moving star wind bubble). 
The sorting effect based on grain-size, shown by~\cite{vanMarle2011} is caused primarily by the fact that the dust grains in that model are already present in the unshocked stellar wind. They tend to decouple from the gas when the latter slows down abruptly as a result of the transition through the wind termination shock. In our present model, the grains are formed in the post-shock region with the same local velocity as the gas. Therefore, there is no sudden velocity change that would cause dust and gas to decouple.
On geometric arguments, we therefore expect the grains, even the larger ones, to remain thightly bound to the mixing zone. 
Making a multiple dust species representation from the single dust bin representation will capture the correct proportionality of small to large dust grains, as it follows the adopted power law with grain size.
The SKIRT dust mixture uses graphite grain properties as specified in~\citet{1993ApJ...402..441L}, again in correspondence to the material density $\rho_p=2$ g\,$\mathrm{cm}^{-3}$ used in MPI-AMRVAC for amorphous carbon grains (which entered the drag force).
One further specifies the photon source, and for the WR 98a system it is appropriate to adopt a (dominant) single stellar source representing the WC9 star (to this end, we used the Potsdam Wolf-Rayet Models\footnote{\url{http://www.astro.physik.uni-potsdam.de/~wrh/PoWR/powrgrid1.html} (using model WC 6-10, accessed on 25 November 2015)} discussed in~\citet{2012A&A...540A.144S}). In accord with this stellar type SED and a typical luminosity of 150000 $L_\odot$, we then launch 1 million photon packages for each of the 50 wavelength bands between 0.006 and 500 $\mu$m that characterize our virtual infrared observations. Dust emission adds another 5 million photon packages accounted for per wavelength band. 

The virtual observation further depends on geometric parameters setting the distance to the object, which we set to 1900 pc, and the relative orientation between the line of sight and the (normal to the) orbital plane. The latter is for WR 98a assumed to be $35^\circ$, but we will vary this inclination from face-on views (zero degree inclination) to edge on ($90^\circ$) views on the expected spiral pattern. Finally, the observing instrument specifics influence the virtual images, as they differ in wavelength sensitivity and resolving power. Our virtual observations will be done for Keck\footnote{{\tt{http://www.keckobservatory.org/}}} (resolution 50 mas at 2.45 $\mu$m), ALMA\footnote{{\tt{http://www.almaobservatory.org/}}} (submillimeter at $\sim 400$ $\mu$m, 6 mas) and near-infrared E-ELT\footnote{{\tt{https://www.eso.org/sci/facilities/eelt/}}} J (6 mas at 1.22 $\mu$m) and K (10 mas at 2.45$\mu$m) specifications.
 For the resolution, our synthetic images have $600\times 600$ pixels covering an area of 1400 AU $\times$ 1400 AU (or 0.00679 pc $\times$ 0.00679 pc), and telescope resolution will be mimicked by convolutions with 20 (Keck), 2 (ALMA and E-ELT J band) or 4 (E-ELT K band) pixel wide point spread functions. In what follows, virtual observations will explore the infrared emission temporal evolution, resolution and relative orientation dependences. We will further contrast infrared views on adiabatic versus cooling scenarios.

\subsubsection{Infrared images}

\begin{figure}
        \vspace{-0.0cm}
        \centering
        \includegraphics[width=0.49\columnwidth]{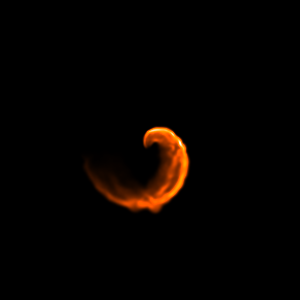}
        \includegraphics[width=0.49\columnwidth]{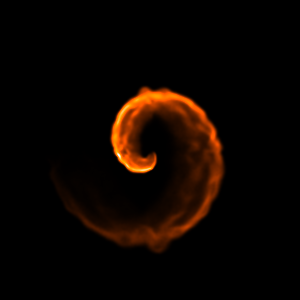}
        \includegraphics[width=0.49\columnwidth]{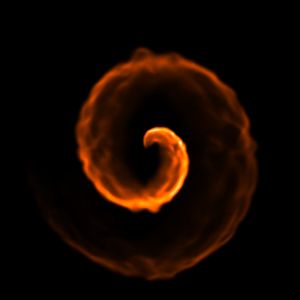}
        \includegraphics[width=0.49\columnwidth]{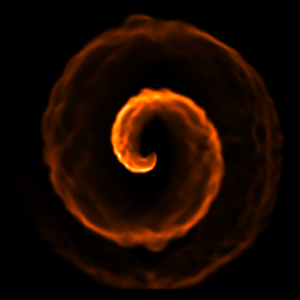}
        \includegraphics[width=0.49\columnwidth]{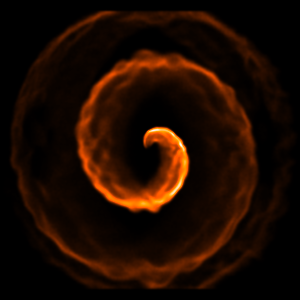}
        \includegraphics[width=0.49\columnwidth]{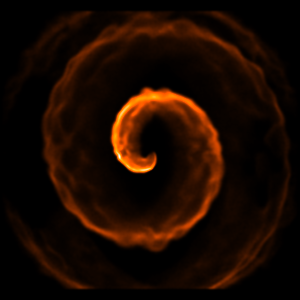}
        \caption{For the run without cooling, synthetic ALMA images at 400 $\mu$m, for times corresponding to 2.4, 2.9, 3.4, 3.9, 4.4 and 4.9 (spanning a time period of 3.9 years). Convolution is done with 2 pixels, from a $600\times 600$ array with an area of $1400\times 1400$ AU$^2$. Our virtual WR 98a system is positioned at 1900 pc, making the field of view $737 {\mathrm{mas}}\times 737 {\mathrm{mas}}$.} \label{synthetic}
\end{figure}

Fig.~\ref{synthetic} contains a sequence of infrared images showing the gradual buildup of the spiral dust pattern within about 4 years time. These are virtual ALMA views on our adiabatic WR 98a simulation, taken at times 2.4, 2.9, 3.4, 3.9, 4.4, and 4.9. The images are for a face-on view, and they serve to show how at ALMA resolution, one can indeed detect the pronounced asymmetry between leading and trailing edge of the spiral pattern, and see the shear-flow induced fine structure evolve with time. In the corresponding 3.9 years between top left and bottom right panel of this figure, the dust clearly gets dragged along into the entire spiral wake carved out by the OB star wind. Moreover, Kelvin-Helmholtz effects lead to undulations growing to a size of order 70 AU within this time period. The dust density decrease along the spiral front leads to infrared emission that correspondingly diminishes.

Assuming a distance to the source of 1900 pc~\citep{vanderHuchtNAR2001}, our model predicts spatially integrated near-infrared fluxes that are roughly compatible with observations. For example, the model predicts a K-band flux of 5 Jy, as compared to observed values of 12.5 Jy~\citep[e.g.,][]{Cutri2003}. At 400 micron, our model predicts an integrated flux of 5 mJ, i.e. three orders of magnitude below the K-band flux. Still, this flux level should be within reach for ALMA observations.

\begin{figure}
        \vspace{-0.0cm}
        \centering
        \includegraphics[width=0.49\columnwidth]{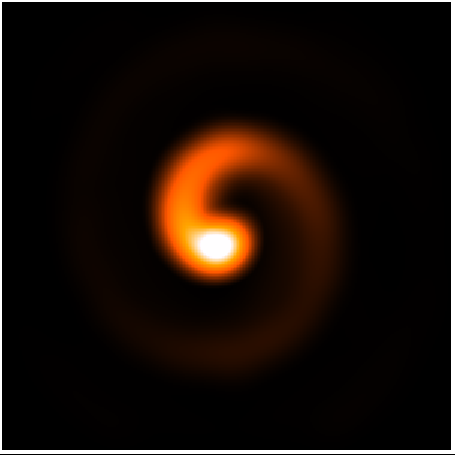}
        \includegraphics[width=0.49\columnwidth]{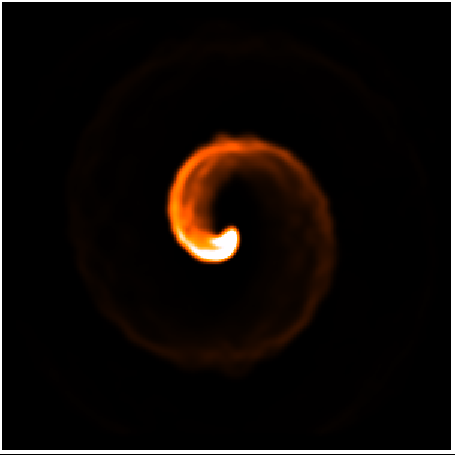}
        \includegraphics[width=0.49\columnwidth]{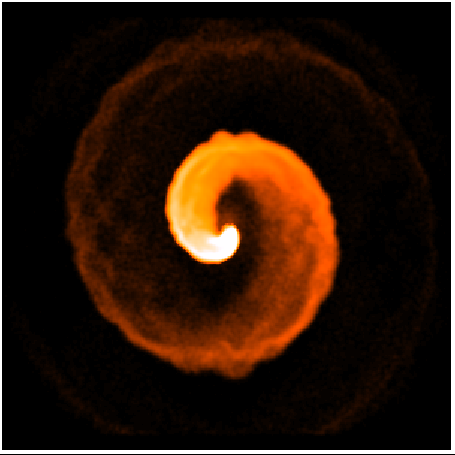}
        \includegraphics[width=0.49\columnwidth]{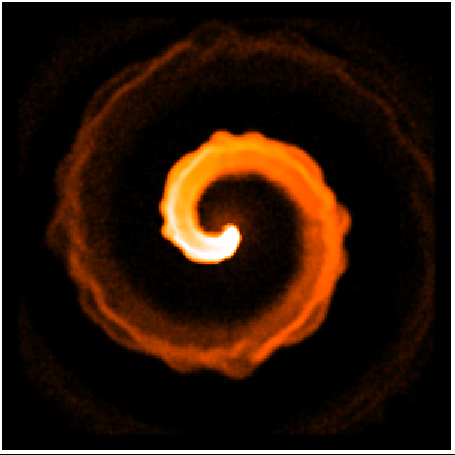}
        \caption{Synthetic images at infrared wavebands, at 2.45$\mu$m (top row) or 1.22$\mu$m (bottom), all for time 4.9 for the run without cooling. Convolution is done with 20, 4 and 2  pixels. These are representative for Keck (top left), E-ELT K band (top right), and E-ELT J band (bottom row). The two bottom images are for inclined (left) versus face-on (right) orientation.} \label{syntheticA}
\end{figure}

Fig.~\ref{syntheticA} demonstrates for the same adiabatic case typical infrared views at shorter wavelengths probed by Keck or E-ELT K or J bands. The top row is for the same time and orientation as the final snapshot from the ALMA sequence in Fig.~\ref{synthetic}, and demonstrates that both resolution and wavelength sensitivity can dramatically influence the virtual images. The (top left) Keck face-on view can no longer distinguish details in the dust density variation, and can not infer the pronounced difference between leading and trailing edge of the dust spiral. In the E-ELT K band view (top right), the smallest grains dominate the emission, and the spiral quickly fades away from detection further down its path as the smaller grains achieve dust temperatures (computed with SKIRT) that drop off more steeply from the center. Indeed, the smallest grains of the 25 dust species have a temperature ranging in [626,4114] K, while the largest species adopt a temperature varying between [245,1831] K. The hottest grains are always found more centrally, and almost all grains have temperatures below the 2200 K sublimation temperature valid for amorphous carbon. The bottom right image shows E-ELT J band imaging, where the same orientation shows a clearly detectable spiral further out. In this waveband, the asymmetry between leading and trailing edge of the spiral is reduced as compared to the longer wavelength ALMA view from Fig.~\ref{synthetic}. Note also that at longer wavelengths, the spiral can be detected all throughout the domain (in the ALMA view, we are also sensitive to emission from the larger grains with their lesser temperature contrast from center to edge). Finally, the bottom left figure for the E-ELT J band contrasts with its bottom right partner in inclination only: the left panel is for a $35^\circ$ inclination angle, versus the $0^\circ$ of the (right panel) face-on view. This latter comparison shows how the viewing angle onto the leading edge of the 3D spiral structure (as shown in Fig.~\ref{gridstuff}, bottom panel) clearly influences the image obtained, with a wider spiral seen in emission when more of the spiral surface area is exposed to the observer. This effect will also influence photometric observations made over several orbital revolutions, as will be discussed in Sect.~\ref{s-photo}.

\begin{figure}
        \vspace{-0.0cm}
        \centering
        \includegraphics[width=0.32\columnwidth]{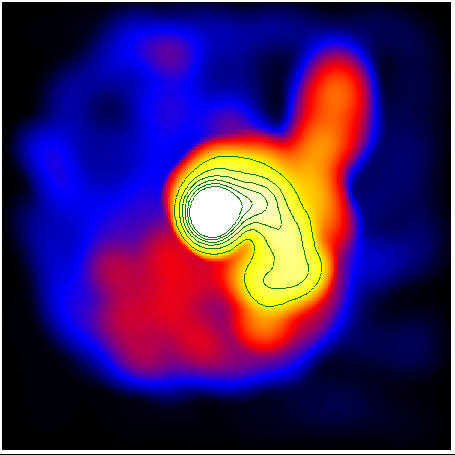}
        \includegraphics[width=0.32\columnwidth]{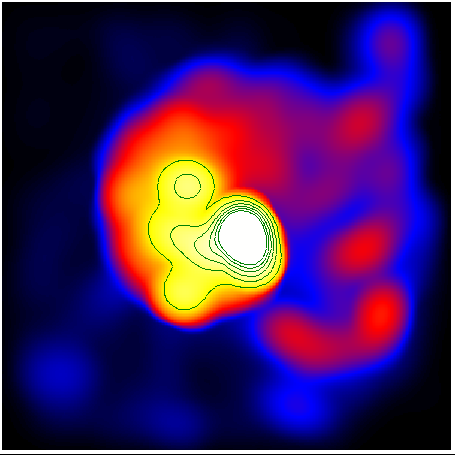}
        \includegraphics[width=0.32\columnwidth]{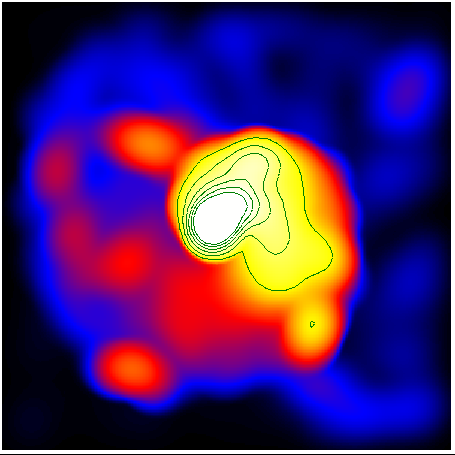}
        \includegraphics[width=0.32\columnwidth]{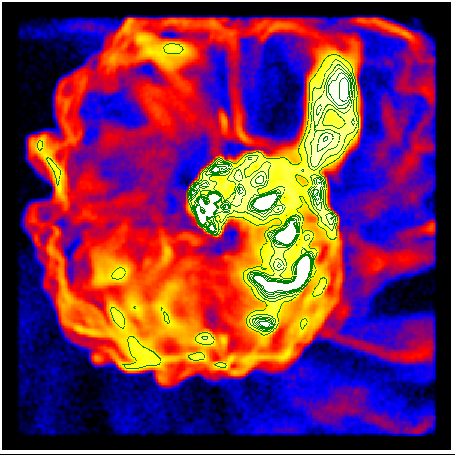}
        \includegraphics[width=0.32\columnwidth]{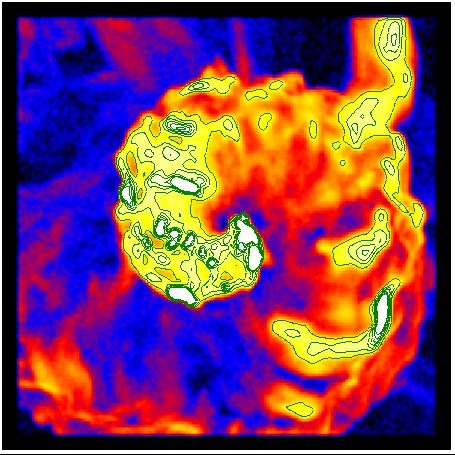}
        \includegraphics[width=0.32\columnwidth]{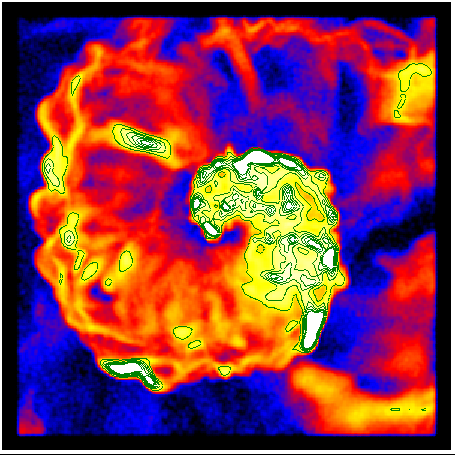}
        \caption{Synthetic images for the run with cooling. Top row for Keck at 2.45$\mu$m, bottom for ALMA at 400$\mu$m, for time 4.5, 5 and 5.5. Convolution is done with 20 versus 2  pixels. We show linearly distributed contours, on top of a logarithmic color image.} \label{syntheticB}
\end{figure}

\begin{figure}
        \vspace{-0.0cm}
        \centering
        \includegraphics[width=0.49\columnwidth]{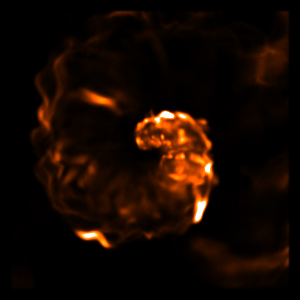}
        \includegraphics[width=0.49\columnwidth]{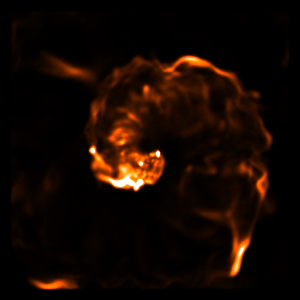}
        \includegraphics[width=0.49\columnwidth]{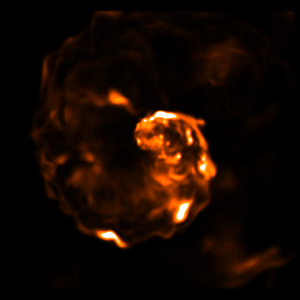}
        \includegraphics[width=0.49\columnwidth]{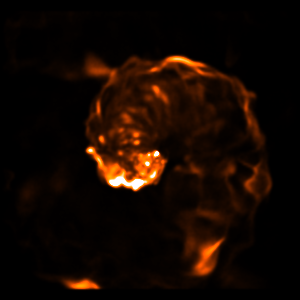}
        \caption{Synthetic images for ALMA on the run with cooling. We compare images for two times $t=5.4$ (left) and 5.9 (right), for face-on (top) versus inclined (bottom) orientation.} \label{syntheticC}
\end{figure}

While both Figs.~\ref{synthetic} and \ref{syntheticA} relate to the simulation without cooling, virtual infrared views on the run with cooling are shown in Figs.~\ref{syntheticB} and \ref{syntheticC}. Figure~\ref{syntheticB} contrasts Keck (top row) with ALMA (bottom row) images at three times in the evolution ($t=$ 4.5, 5, 5.5). This demonstrates that resolution is once more key to clearly distinguish the very fragmented mixing zone where dust now gets concentrated in a more spherical (rather than clearly spiral) fashion, dominated by local dust clumps and filaments (as shown in Fig.~\ref{3dWRClumps}). Figure~\ref{syntheticB} shows in each panel how a linear color scale (shown with contours) no longer yields a clearly detectable rotating spiral pattern, as outward moving large localized clumps may seem unaligned with the underlying orbital motion. The color scale in Fig.~\ref{syntheticB} plots the same data in a logarithmic scale (as was also adopted for all previous figures), where only in the ALMA views, one can still find evidence for a rotating spiral pattern. Again, this longer wavelength view shows grain emission from both the large and small grains. Note that the possible failure to detect a clear spiral pattern in infrared views may have significant consequences for interpreting actual infrared evolutions, where binarity is easily linked with clear rotating spirals: our findings suggest that more erratically varying dusty WR environments could be binary systems where radiative cooling effects induce the variability. Disentangling episodic dust makers from binary configurations using only infrared variability is thus poorly constrained.

Finally, Fig.~\ref{syntheticC} contrasts virtual ALMA views at times 5.4 (left column) and time 5.9 (right column) for the case with cooling (using the same representation as for the adiabatic case in Fig.~\ref{synthetic}). The difference between top and bottom row is purely in orientation, where the top view is for the face-on line of sight, while the bottom has an inclination of $35^\circ$. Unlike the adiabatic case from Fig.~\ref{syntheticA}, where the overall orientation with respect to the spiral obviously mattered, the case with cooling shows little variation with viewing angle. This is in line with the more volume-filling, spherical redistribution of the mixing zone, as well as with the fact that the views are always dominated by the densest clumps and filaments. This effect will return in virtual photometric observations, which we discuss in the next paragraph.

\subsubsection{Virtual photometric observations}\label{s-photo}

\begin{figure}
        \vspace{-0.0cm}
        \centering
        \includegraphics[width=\columnwidth]{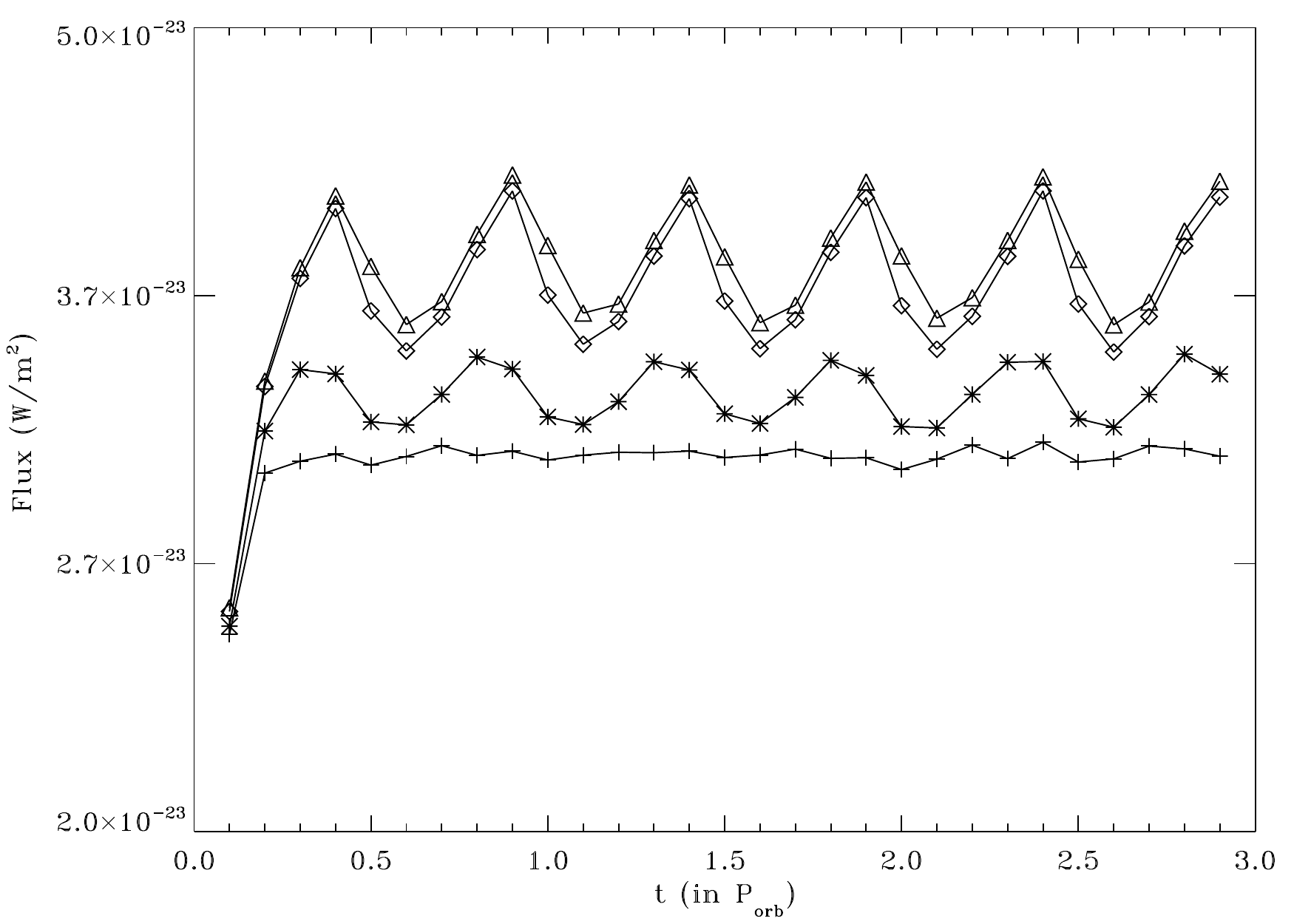}
        \includegraphics[width=\columnwidth]{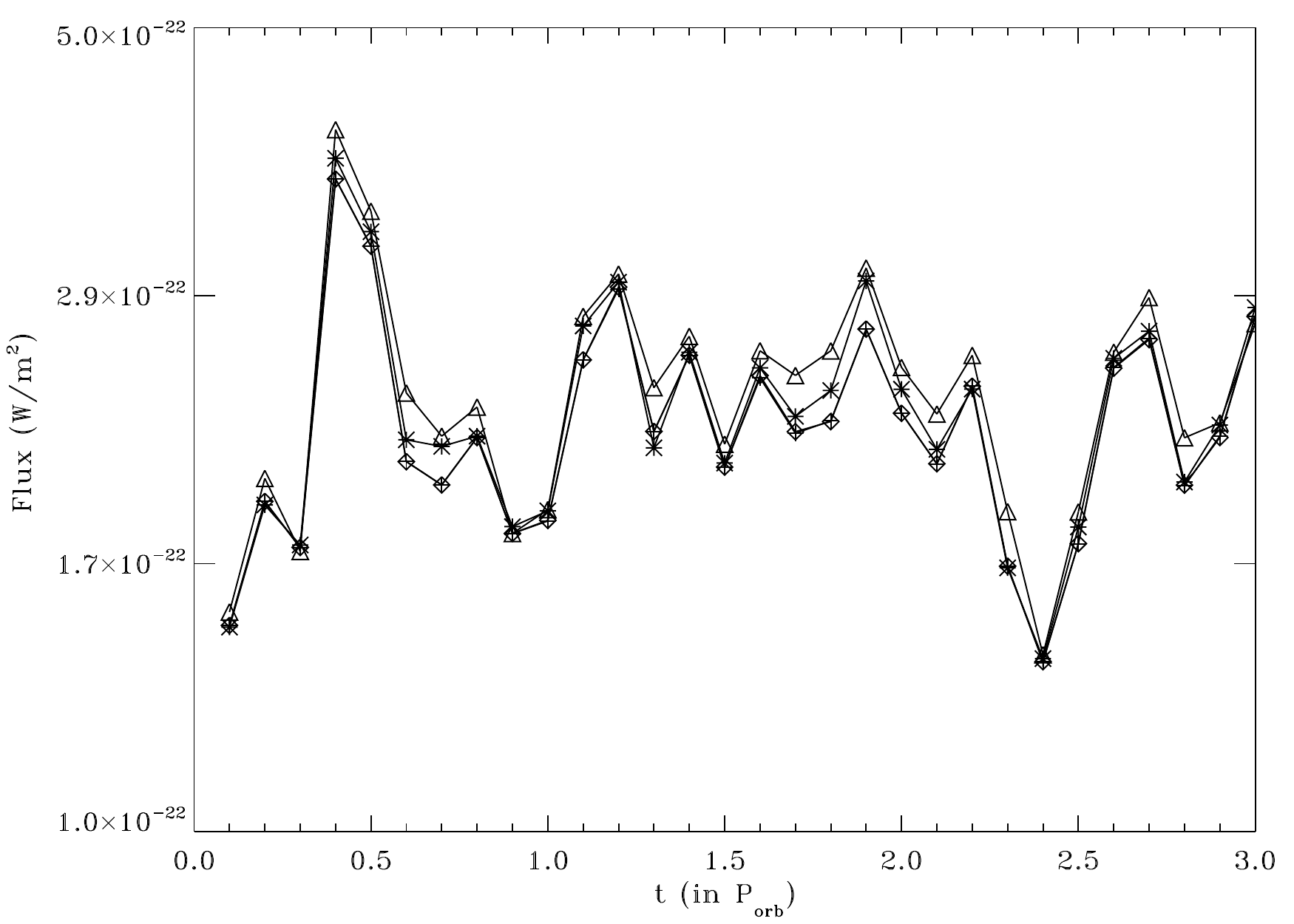}
        \caption{Photometric observations over three orbits (or 4.6 years). Top row is for the case without cooling, bottom is with cooling. Each panel shows 4 different inclinations, varying from face-on ($+$), to $35^\circ$ ($*$), $70^\circ$ (diamond), edge-on (triangles).} \label{photometric}
\end{figure}

The fact that we obtained virtual infrared views over areas corresponding to 1300 AU $\times$ 1300 AU in the orbital plane, and this for 3 complete orbits, also allows us to mimic photometric observations made over several years. We aim here to illustrate how inclination effects influence the observed photometric variability. We obtain a lightcurve as follows: for the specific wavelength of 1.22 $\mu$m corresponding to the E-ELT J band, we first generate a full resolution (unconvolved) sequence of 29 virtual infrared images for all times $t=2.0, \, 2.1, \,2.2, \ldots 4.9$ (for the adiabatic case). Each image is then turned into a single photometric flux quantification, by summing over all flux values within a preset, 25 pixel aperture radius centered on the image. We repeat this process for 4 inclinations, and thereby obtain the top panel of Fig.~\ref{photometric}, where we compare face-on ($0^\circ$ inclination, shown with crosses), $35^\circ$ inclination (asterisk symbols), $70^\circ$ (diamonds) and side (triangles) views. For WR 98a, actual infrared observations (imaging and photometry) led to infer a $35^\circ$ inclination between the orbital plane and the LOS. Our virtual photometric observation indeed shows that at sufficiently large inclinations with respect to the orbital plane, the orbital variability will clearly leave its imprint on the lightcurve. The effect disappears for pure face-on views, as expected (the precise orientation of the spiral in the orbital plane becomes irrelevant in photometry). The changes demonstrated due to orientation effects as shown in Fig.~\ref{syntheticA} lead to flux changes which are most pronounced for a sideway view: this is understood from the spiral asymmetry between leading and trailing edge, together with the orbit-related phase-dependence. The flux change at a $35^\circ$ orientation is similar in trend to the one reported by observations in \citet{1995MNRAS.275..889W,1999ApJ...525L..97M}. 
The actual flux values we obtain here are not matching in order of magnitude, since actual photometric surveys collect photons over a long time interval, over a much more extended area of the sky than what we can extract from our limited 3D domain (here we only take a central region, as we do not want to be affected by the build-up of the spiral pattern over time).

The bottom panel of Fig.~\ref{photometric} does exactly the same virtual photometry, but this time from 30 frames in the 3 orbits of the cooling case, with dust. As more dust is present in this case, the flux levels are up by one order of magnitude with respect to the adiabatic case. More importantly, the lightcurve can hardly distinguish between the rather different orientations (face-on to edge-on), once more confirming that with effective cooling, a much more clumpy dust distribution gets established due to the prevalence of thermal instabilities. The mixing happens in a more spherically symmetric, erratic fashion, and orientation effects in the infrared images (as shown in Fig.~\ref{syntheticC}) largely disappear, to become undetectable in photometry alone. For the case of WR 98a, we are thus led to infer that in this binary WC9+OB system, radiative cooling effects are likely to be inefficient.

\section{Conclusions}

We presented a 3D gas plus dust evolution model for WR 98a, taking into account knowledge on the stellar winds and orbital parameters of the system, as well as observationally well-constrained aspects of dust formation. Using block-adaptive capabilities, we covered several orbital revolutions (i.e. 7.5 up to 9.3 years in total) on scales large enough to allow direct confrontation with the latest infrared observations (Keck and ALMA), or for predicting near-infrared views at future E-ELT specifications. We showed that in adiabatic evolutions, the large-scale mixing zone automatically organizes into a spiral pattern, which is only slightly disturbed by shear-flow related instabilities, especially on the leading flank. A clear asymmetry between the leading and trailing part of the spiral develops, with more dust accumulated in the narrower, more compressed leading part. A more wide and diffuse mixing layer is found at the trailing side, and this asymmetry must be probable at ALMA resolutions (i.e. below 10 mas) but fails to be detected with Keck specifications (50 mas resolution). We also predict that fine structure due to Kelvin-Helmholtz development can be seen as ripples on the main spiral pattern, and each undulation can be followed to grow to a size of order 70 AU within several years. Our virtual photometric observations confirm that under the $35^\circ$ relative orientation between the normal to the orbital plane and the LOS, we can detect the orbital variation as a clear modulation of the lightcurve. On the other hand, we also simulated a scenario for WR 98a where optically thin radiative cooling is taken along. This scenario is dominated by a much more erratic density structuring on the large scale, with an enhanced mixing volume and dust creation zone. This is already seen at scales of the orbital seperation, as the WCR around the OB star has an enlarged semi-opening angle. Thermal instabilities induce very large density contrasts compared to an adiabatic case, and much more dust gets created in the localized high density clumps. The virtual infrared appearance of a case with effective cooling does not show a clear pinwheel structure, and inclination effects hardly matter in photometric evolutions. 

Our findings imply that the WR 98a environment is not influenced much by strong radiative cooling. At the same time, binary systems with effective cooling may not show a pronounced rotating pinwheel structure, and may appear more episodic in their dust evolution. Future modeling efforts may try to incorporate details in the stellar wind prescriptions itself, as at least for single WR stars, clumpy winds are the only way to explain how dust can form and survive in the extreme temperature conditions and luminosities of WR stars and their winds. It will then also be of interest to make virtual images for other interferometry instruments, like VLTI-MATISSE\footnote{\url{http://www.matisse.oca.eu}} operating in the wavelength range $2.8-13\,\mu\mathrm{m}$.

Our model shows distinct differences in the variability of the colliding wind shock front between cases with weak to strong radiative cooling. When we quantify the mean temperature in the mixing zone (where $\theta_{OB}\theta_{WR}>500$) between our inner and outer dust formation distances $D_{in}$ and $D_{out}$ for the case without cooling, we obtain a value of $1.5 \times 10^6$ K (or 129 eV) for the 0.004 volume fraction it occupies, which we reported has a mean density of $10^{-20} \,\mathrm{g}/\mathrm{cm}^3$. For the case with cooling, we already noted that the volume fraction increases to 0.023, but the mean temperature has dropped to values well below $10^5$ K (or 8.6 eV), while a mean density in the mixing zone is $10^{-18} \,\mathrm{g}/\mathrm{cm}^3$ (an order of magnitude higher than the already reported mean density for the entire domain). This hot thermal plasma would emit radiation with the essential proportionality $(\rho/m_p)^2 \sqrt{T}$ (with $T$ in eV and number densities in inverse cubic centimeter), which is equal to $4\times 10^8$ for the non-cooling case, where it is emitted quasi-steadily from a small volume. In the radiative case, we find a value of about $10^{12}$, and this from a highly time-variable larger volume. The fact that WR98a appears to have only a weak X-ray flux is thus fully consistent with our finding that it realizes conditions without strong radiative cooling. At the same time, a lack of X-rays can also be explained by them being absorbed, as e.g. is the case in the $\eta$ Carinae system, where full 5.5 year orbit X-ray monitoring showed distinct eclipsing occuring in 2-10 keV emission~\citep{Hamaguchi2014}.

Finally, we may speculate on how similar modeling efforts for systems like WR 104, WR 140, or WR 122 would lead to further insights in the behavior of dusty pinwheel nebula. From the large-scale modeling perspective, the main parameters determining the overall interaction morphology will be the stellar parameters and their wind mass loss rates and speeds, in turn fixing the momentum flux ratio. Our model supplements the findings from earlier models that, together with the separation distance between the binary companions, this ratio is very important for the large scale stability of the spiral. The orbital parameters also play a decisive role, since the orbit (hence separation) determines whether the winds are already at full terminal speed or need to consider near-star acceleration (and perhaps, wind variability) for modeling the wind collision zone correctly. Naturally, we can expect eccentric systems to have additional orbit-related variability, and it is this fact that is often invoked for explaining episodic dust creation in systems like WR 140. We argue that even a non-eccentric system, where the conditions are favorable to optically thin radiative loss effects like those included in our second model, would likewise appear episodic, in the sense that the clumpiness we found for these systems is seen to rotate along during the orbit. In that sense, we stated that disentangling episodic dust makers from eccentric to non-eccentric configurations is poorly constrained, if the orbital parameters are not precisely known.

\section*{Acknowledgements}

This research was supported by the Interuniversity Attraction Poles
Programme (initiated by the Belgian Science Policy Office, IAP P7/08 CHARM) and by the KU Leuven GOA/2015-014. Computational resources and services used in this work were provided by the VSC (Flemish Supercomputer Center) funded by the
Hercules foundation and the Flemish government, department EWI, as well as PRACE resources (project pr87di) on SuperMUC at Garching. We thank the first referee for constructive comments and suggestions.




\bibliographystyle{mnras}



\appendix

\section{Dust formation model}
\label{dustFormModAppendix}

In this appendix, we explain our choice for fixing the semi-empirical parameters $\phi$ and dust formation distances $D_{in}$ to $D_{out}$. We first argue that a semi-empirical parametrization of the dust insertion is necessary, as combining grain growth models with large-scale simulations is as yet impracticle.
Indeed, to be able to put dust formation in simulations in a physically sound fashion, not only do we need to know \textit{where} dust forms, but equally important is knowing \textit{at which rate} dust forms at those locations. Dust formation consists of two distinct steps: nucleation and growth. Dust nucleation is a chemical process that forms large molecules and nanocrystals ($\sim$ 4 \AA) from which dust grains can start to grow. Dust growth is mainly a physical process which allows the dust precursors to grow further through accretion and collisions. Understanding how exactly these processes work around dust forming WR stars turns out to be a difficult question to answer from a theoretical point of view. The chemical growth models used in \citet{2000A&A...357..572C} and \citet{2002ASPC..260..223L} are unable to produce significant amounts of dust precursors in WR environments, unless extremely high densities are achieved. \citet{1998MNRAS.295..109Z} investigated dust grain growth around several WR stars, starting from 4 $\AA$ dust precursors. In these models grains achieve growth up to sizes between 10-20 nm, and the synthetic spectra of the resulting grains provide a good match with the observed infrared spectra of the WR stars. Observations seem to find (significantly) larger grains. \citet{2003ApJ...596.1295M} find that the characteristic size of the dust grains in WR 140 is around 70 nm. For WR 112 \citet{2002ApJ...565L..59M} find much larger grains, around $s = 500$ nm, which they indicate is significantly larger than theoretical limits. Micron sized grains are supported by the analysis of \citet{2001ApJ...550L.207C} for several WC stars (WR 104, WR 112, and WR 118). \citet{2001A&A...379..229Y} modelled the 2.13 $\mu$m speckle interferometric observations of WR 118 with a radiative transfer model, and they report that the SED can either be fitted with a mix of small and large grains ($s=50$ nm and $s=380$ nm), or by using a particle size distribution $n(s) \propto s^{-3}$ with $s$ between 5 nm and 600 nm. The underestimation of grain growth by the theoretical models \citep{1998MNRAS.295..109Z} can (partly) be attributed to local underestimations of the density: clumps of high density material may be formed in the WCR. The density of these clumps can be further increased due to local or large scale thermal instabilities in the WCR. Taking these density increases into account when simulating the large scale structure of the binary is practically impossible as small scale structures tend to be unresolved, especially those due to thermal instabilities, which will push dense structures to smaller scales and even higher densities. Due to these reasons, we can not reasonably hope to apply a dust growth model on the local values in the large scale simulations we would like to perform.

While the \textit{local} dust formation rate which we just discussed is difficult to model, the \textit{global} dust formation rate $\dot{M}_d$ is difficult to constrain from observations because often it is difficult to infer the total mass-loss of the system. Table \ref{convertionRate} gives an overview of several dust conversion fractions $\xi = \dot{M}_d / \dot{M}_{WR}$ estimated from observations, and clearly a large range of values exist for $\xi$. The dust growth models of \citet{1998MNRAS.295..109Z}, which take $\dot{M}_{WR} = 8\times 10^{-8}$ $M_\odot$ yr$^{-1}$ for all objects, predict lower values of $\xi$ for two objects that are covered in other studies: 0.66 \% for WR 104 and 0.04 \% for WR 112.

\begin{figure}
    	\vspace{-0.0cm}
	\centering
	\includegraphics[width=\columnwidth]{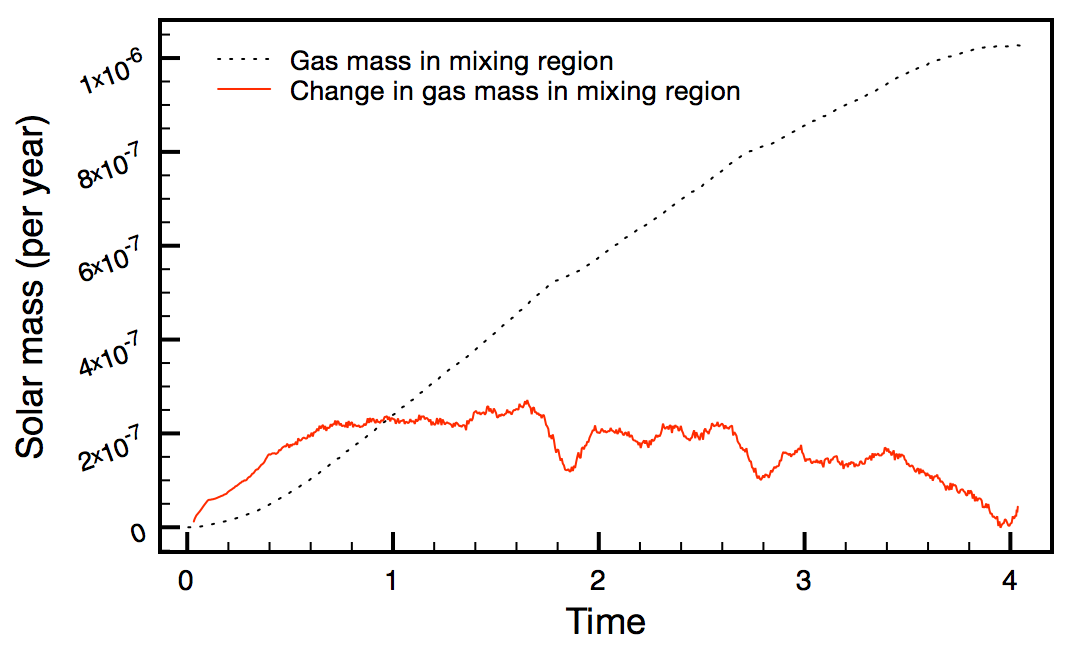}
	\caption{Evolution of the total gas mass in the mixing layer $M_{mix}$ (in $M_\odot$) and the change of mass in the mixing layer $\dot{M}_{mix}$ (units of $M_\odot$ yr$^{-1}$) in a 3D simulation without dust insertion (and no cooling). Time is given in orbital periods. This full dynamical simulation in turn allows to estimate the size of the dust forming region and the local dust growth, which are difficult to obtain from observations or from dust growth theory.}
	\label{mixLayerEvolve}
\end{figure}

Given all these considerations, we now relate our semi-empirical parametrization to these findings, fixing the parameters $\phi$, $D_{in}$ and $D_{out}$.
The latter two are somewhat constrained by observation, especially the distance from which dust can form can be argued to  be $D_{in} = 20.3$ AU, as explained in section~\ref{dustFormWR98a}. The local dust formation rate $\phi$ is more difficult to obtain. However, we can predict values for these quantities from our dynamical simulations in the following way: first we run a 3D simulation without any dust insertion. In this simulation, the tracers in the two stellar outflows track the region in which mixing takes place. The value of each tracer fluid ranges between $[0,100]$. We define the region of significant mixing to be where  $500 < \theta_{WR}\theta_{OB}$, meaning that (1) the region must contain material from both stars, (2) in a region where a local tracer identifies 100 for one of the stars, the other star matter must contribute at least 5\%. In practice this means that in the mixing region the dense WR material is enriched with hydrogen. This simulation without any dust formation allows us to calculate the total mass of gas in this mixing region as a function of time $M_{mix}(t)$. 
The same simulation also then quantifies $\dot{M}_{mix}$, i.e. the change in gas mass added to the mixing region. A quantification of $M_{mix}(t)$ and $\dot{M}_{mix}$ from this simulation is shown in Fig.~\ref{mixLayerEvolve}. It turns out that we find
$\dot{M}_{mix} \approx 2 \times 10^{-7} M_\odot \text{yr}^{-1}$, or 4\% of the WR mass-loss.
The value of $\dot{M}_{WR}$ is deduced from observations of WR 98a as mentioned in section \ref{physSetupwr98a}. 

From the literature on WR objects known to produce dust, one also finds a fairly wide range of 
(see table \ref{convertionRate}) the gas-to-dust conversion fraction $\xi = \dot{M}_d / \dot{M}_{WR}$. 
Since the range varies strongly in the literature, it is difficult to directly find a value for WR 98a. If we take an average value of $\xi  = 0.2 \%$ (average among the various objects mentioned in table \ref{convertionRate}), we can relate the total dust formation rate $\dot{M}_{d}$ into a percentage of material from the mixing layer that is converted to dust, $\chi$:
\begin{align}
\dot{M}_d = \xi \dot{M}_{WR} &= 0.002\times (5\times 10^{-6} M_\odot \text{yr}^{-1}) \\
&= 10^{-8} M_\odot \text{yr}^{-1} = \chi \dot{M}_{mix}\\
\Rightarrow \chi &= \frac{10^{-8} M_\odot \text{yr}^{-1}}{\dot{M}_{mix}}.
\end{align}
Since our dynamical simulation in 3D quantifies $\dot{M}_{mix} \approx 2 \times 10^{-7} M_\odot \text{yr}^{-1}$ (see Fig.~\ref{mixLayerEvolve}), we would deduce a dust conversion fraction in the mixing layer of $\chi = 0.05$.

To get the local dust conversion rate $\phi$, we need to know how much time the gas in the mixing region spends in the region where dust can form: between $D_{in}$ and $D_{out}$. We call this time 
\begin{equation}
t_{cross} = \frac{D_{out} - D_{in}}{v_{mix}} = \frac{\Delta D}{v_{mix}} ,
\end{equation}
with $v_{mix}$ the local radial velocity. If we assume that $v_{mix} \approx v_{WR} = 900$ km s$^{-1}$, we can derive $t_{cross}$ and $\Delta D$ from
\begin{align}
\dot{M}_d &=  10^{-8} M_\odot \text{yr}^{-1} = \frac{\chi M_{mix}}{t_{cross}} \\
&=\frac{\chi M_{mix}}{t - t_{0}} 
\end{align}
with $t_0$ the time at which the mixing region starts to form. Using the simulated data for $M_{mix}(t)$ as shown in Fig.~\ref{mixLayerEvolve}, we find that $t_{cross} \approx 0.685 P_{orb} = 3.2 \times 10^{7}$ s, from which we get the local dust formation rate $\phi$ and the width $\Delta D$ of the region where dust is formed:
\begin{align}
\phi &= \frac{\chi}{t_{cross}} \approx 0.0763\\
\text{and} \qquad \Delta D &= v_{mix} t_{cross} \approx 192.5 AU = 47.4\,a. 
\end{align}
If we then take the distance from which dust can form to be $D_{in} = 20.3$ AU, we obtain $D_{out} = 212.8$ AU, both of which are realistic values given the observations we discussed.

\begin{table}
\caption{Estimated values of the dust conversion rate $\xi$.}
\label{convertionRate}
\centering
\begin{threeparttable}
\begin{tabular}{l c c c}    
\hline\hline 
$\xi$						& object	& reference & note	\\
\hline 
2.67 $\%$					& WR 104	& (1)	& radiative model\\
6 $\%$					& WR 112	& (2)	& observation \\
7 $\times 10^{-4}$ $\%$ - 7 $\%$  & 17 objects & (3) & grain growth models\\
0.06 \%					& WR 139 & (4) &observation \\
0.2 $\%$					& WR 140	& (5)	& observation \\
\hline        
\end{tabular}
 \begin{tablenotes}    
\item[] (1): \citet{2004MNRAS.350..565H} (2): \citet{2002ApJ...565L..59M} (3): \citet{1998MNRAS.295..109Z}  (4): \citet{1992A&A...261..503H} \& \citet{2001MNRAS.324..156W} (5): \citet{1987A&A...182...91W} \& \citet{1990MNRAS.243..662W} 
     \end{tablenotes}
     \vspace*{0.2cm}
  \end{threeparttable}
\end{table}

\section{3D Simulations meet 3D printing}
\label{s:3Dprint}

\begin{figure}
    	\vspace{-0.0cm}
	\centering
	\includegraphics[width=\columnwidth]{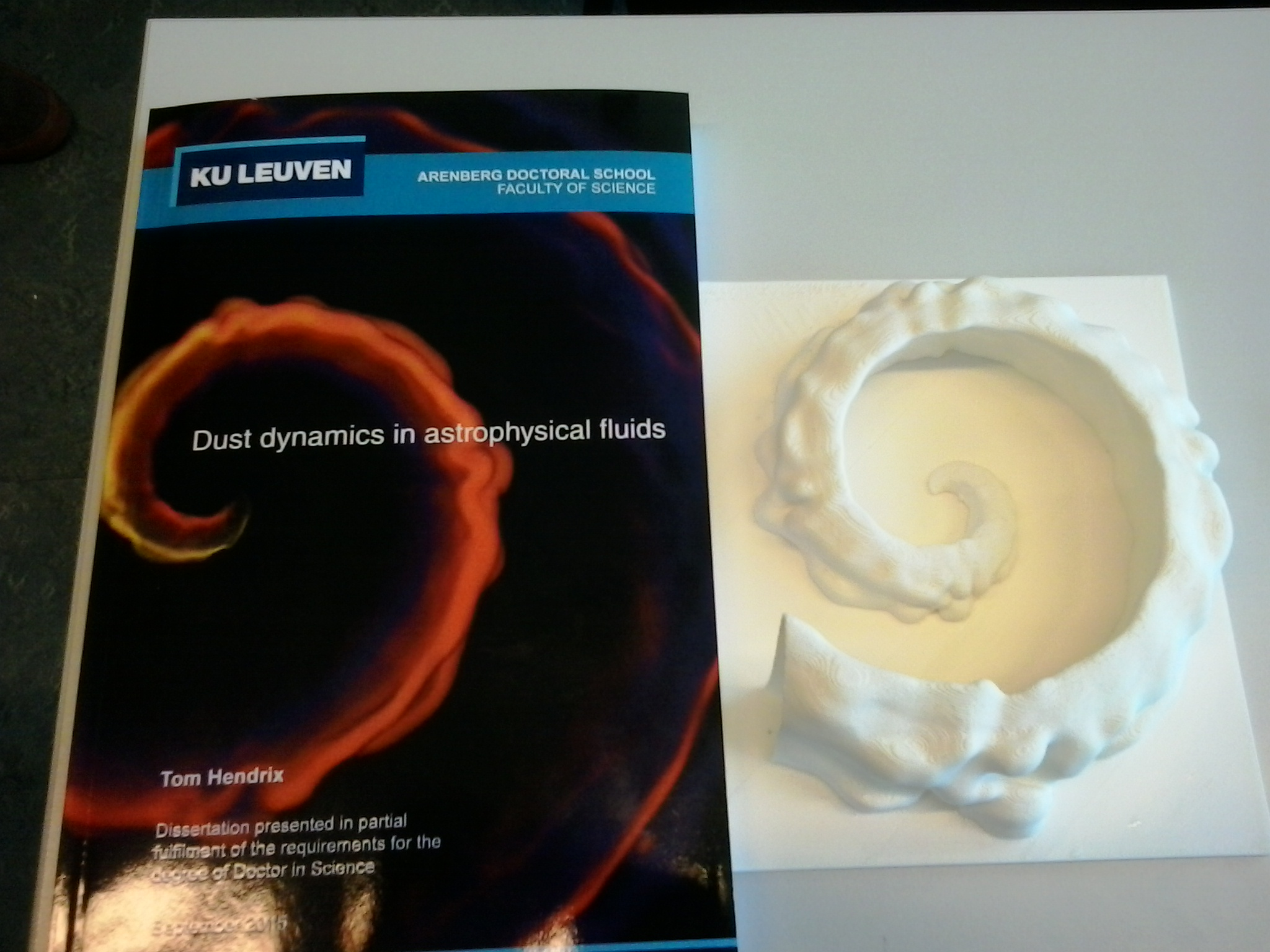}
	\caption{A 3D printout of the hydro gas plus dust model of WR 98a. Printed is an isosurface of the mixing ratio.}
	\label{f:3Dprint}
\end{figure}

The MPI-AMRVAC code~\citep{2014ApJS..214....4P} has many options for output and data analysis, during runtime and/or in post-processing. Most of the figures in this paper used its capacity to convert its native {\tt .dat} format for data snapshots to {\tt .vtu} format, which can directly be used for animations using {\em VisIt}, or {\it Paraview}. In order to create a 3D print from a hydrodynamical simulation, one starts by defining a surface area, using the iso-surface option available in these programs. Once a representative surface has been obtained, this can then be saved in {\tt .stl} format. The {\tt .stl} file needs to be checked for flaws (e.g., the printout consisting of seperate non-connected regions, which would obviously fall apart) and, if necessary, repaired. The final result can then be uploaded to the 3-D printer. A result is shown in Fig.~\ref{f:3Dprint}.

\bsp	
\label{lastpage}
\end{document}